\documentclass[conference]{IEEEtran}
\IEEEoverridecommandlockouts

\usepackage{cite}
\usepackage{amsmath,amssymb,amsfonts}
\usepackage{algorithmic}
\usepackage{graphicx}
\usepackage{textcomp}
\usepackage{xcolor}
\usepackage{tabularx}
\usepackage{colortbl}
\usepackage{caption}
\usepackage{array}
\usepackage{stfloats}
\usepackage{url}
\usepackage{verbatim}
\usepackage{subcaption}
\usepackage{longtable}
\usepackage{float}
\usepackage{tabularray}
\usepackage[utf8]{inputenc}
\usepackage[T1]{fontenc}

\def\BibTeX{{\rm B\kern-.05em{\sc i\kern-.025em b}\kern-.08em
    T\kern-.1667em\lower.7ex\hbox{E}\kern-.125emX}}
\begin{document}

\title{QuLore: An Adaptive Security Framework to Extend Quantum-Safe Communications to Real-World Networks\thanks{This manuscript has been submitted for possible publication.}}

\author{\IEEEauthorblockN{Ane Sanz}
\IEEEauthorblockA{\textit{\textsuperscript{a}Dept. of Communications Engineering} \\
\textit{\textsuperscript{b}EHU Quantum Center} \\
\textit{University of the Basque Country}\\
Bilbao, Spain \\
ane.sanz@ehu.eus}
\and
\IEEEauthorblockN{Eire Salegi}
\IEEEauthorblockA{\textit{\textsuperscript{a}Dept. of Communications Engineering} \\
\textit{University of the Basque Country}\\
Bilbao, Spain \\
eire.salegi@ehu.eus}
\and
\IEEEauthorblockN{Asier Atutxa}
\IEEEauthorblockA{\textit{\textsuperscript{a}Dept. of Communications Engineering} \\
\textit{\textsuperscript{b}EHU Quantum Center} \\
\textit{University of the Basque Country}\\
Bilbao, Spain \\
ane.sanz@ehu.eus}
\and
\IEEEauthorblockN{David Franco}
\IEEEauthorblockA{\textit{\textsuperscript{a}Dept. of Communications Engineering} \\
\textit{University of the Basque Country}\\
Bilbao, Spain \\
david.franco@ehu.eus}
\and
\IEEEauthorblockN{Jasone Astorga}
\IEEEauthorblockA{\textit{\textsuperscript{a}Dept. of Communications Engineering} \\
\textit{\textsuperscript{b}EHU Quantum Center} \\
\textit{University of the Basque Country}\\
Bilbao, Spain \\
jasone.astorga@ehu.eus}
\and
\IEEEauthorblockN{Eduardo Jacob}
\IEEEauthorblockA{\textit{\textsuperscript{a}Dept. of Communications Engineering} \\
\textit{\textsuperscript{b}EHU Quantum Center} \\
\textit{University of the Basque Country}\\
Bilbao, Spain \\
eduardo.jacob@ehu.eus}
}

\maketitle

\begin{abstract}
The advent of quantum computing threatens classical cryptographic mechanisms, demanding new strategies for securing communication networks. Since real-world networks cannot be fully Quantum Key Distribution (QKD)-enabled due to infrastructure constraints, practical security solutions must support hybrid operation. This paper presents an adaptive security framework that enables quantum-safe communications across real-world heterogeneous networks by combining QKD and Post-Quantum Cryptography (PQC). Building upon a hierarchical key management architecture with Virtual Key Management Systems (vKMS) and a centralized Quantum Security Controller (QuSeC), the framework dynamically assigns security levels based on node capabilities. By transitioning between pure QKD, hybrid, and PQC modes, it ensures end-to-end quantum-safe protection regardless of the underlying node capabilities. The framework has been implemented and validated on a Kubernetes-based containerized testbed, demonstrating robust operation and performance across all scenarios. Results highlight its potential to support the gradual integration of quantum-safe technologies into existing infrastructures, paving the way toward fully quantum-safe communication networks.

\end{abstract}


\section{Introduction}
The development of quantum computing introduces a paradigm shift in the field of cybersecurity, with direct implications for quantum communication and quantum cryptography. While quantum computers promise revolutionary advancements in several fields, they also pose a significant threat to classical cryptographic algorithms. In particular, widely used public-key algorithms such as RSA and ECC are considered vulnerable to quantum attacks, potentially compromising the asymmetric primitives that underpin key exchange and digital signatures in modern communication systems~\cite{easttom2022quantum}. This renders current mechanisms for establishing and managing cryptographic keys insecure. 

To address these threats, two primary approaches have been developed for achieving quantum-safe key establishment. The first approach is Post-Quantum Cryptography (PQC), which replaces classical public-key algorithms with alternatives based on hard mathematical problems believed to resist quantum attacks. PQC algorithms can be implemented using existing classical hardware infrastructure, making them easily deployable across current networks. However, their security ultimately depends on computational assumptions \cite{jenefa2023pqc}.

The second approach is Quantum Key Distribution (QKD), which avoids public-key cryptography entirely by generating symmetric keys between two legitimate parties using the principles of quantum mechanics. Unlike PQC, QKD provides Information-Theoretic Security (ITS), guaranteed by physical laws rather than computational hardness assumptions \cite{nadlinger2022experimental}. 

However, several challenges hinder its widespread adoption within current communication infrastructures. First, QKD requires specialized quantum hardware, which increases deployment cost and complexity \cite{diamanti2016practical}. Second, the achievable distance is limited by the lack of quantum repeaters, restricting QKD to metropolitan-scale links. Third, extending QKD beyond point-to-point connections often requires trusted relay nodes, which may weaken the security guarantees by introducing additional points of vulnerability. Fourth, many QKD systems require dedicated dark fibers for quantum signal transmission, complicating integration with existing optical infrastructures and limiting scalability. 

\begin{figure}[h!]
\centering
    \includegraphics[width=0.8\columnwidth]{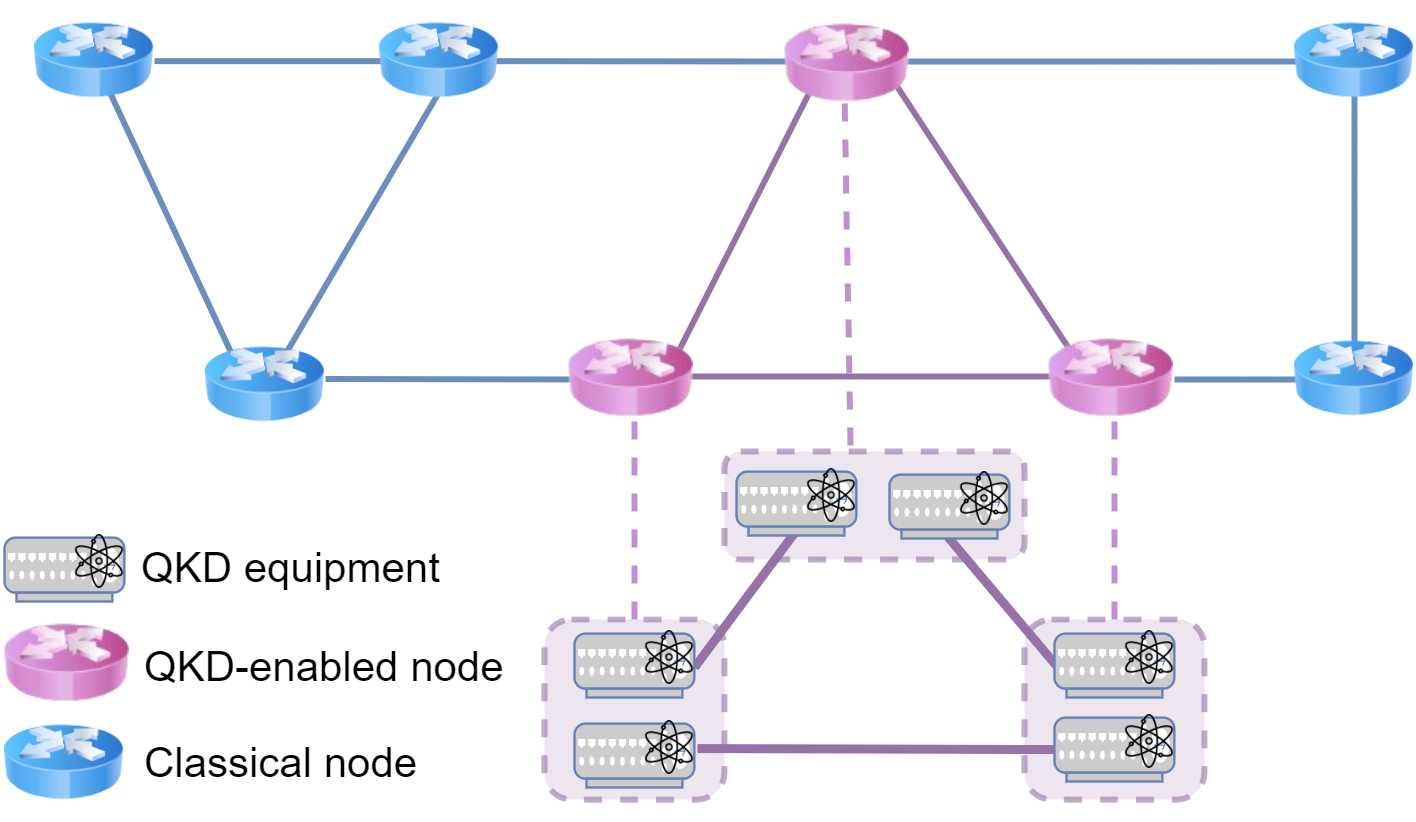}
\caption{A representation of a heterogeneous network comprised of QKD-enabled and classical nodes.}
\label{fig:heterogeneousNetwork}
\end{figure}

Although a fully QKD-enabled network represents the ideal quantum-safe communication paradigm, current constraints make such deployments impractical today. Instead, real-world networks are usually heterogeneous, comprising both QKD-enabled and classical nodes that rely on PQC, as depicted in Figure~\ref{fig:heterogeneousNetwork}. Each node is defined as a site that may include one or more pieces of network equipment, quantum equipment, or other related infrastructure. Hybrid architectures that strategically integrate both approaches offer a realistic path forward: they enable the gradual incorporation of QKD while ensuring continuity of service through PQC. This heterogeneity, however, introduces new challenges for providing end-to-end (E2E) quantum-safe communication across mixed networks.

Our previous work on scalable QKD architectures \cite{sanz2025toward} introduced a hierarchical Key Management System (KMS) with Virtual KMSs (vKMS) and a centralized Quantum Security Controller (QuSeC) for efficient KMS discovery and multi-hop relay path computation. Building on that work, this paper extends the model to heterogeneous networks. While that earlier solution improved scalability with compliance with ETSI specifications, it assumed universal QKD capability, limiting its applicability in transitional or hybrid environments.

To address this gap, we propose an adaptive security framework that unifies QKD and PQC under centralized management. The framework introduces a layered architecture with vKMSs and a QuSeC that dynamically assigns one of four security levels, ranging from pure QKD to pure PQC, with hybrid strategies that leverage Key Derivation Functions (KDFs). This ensures that each communication session receives the strongest feasible quantum-safe protection, regardless of node capabilities. Therefore, the main contributions of this work are the following:

\begin{enumerate}
    \item \textbf{Unified Architecture:} An adaptive security framework that extends quantum security to heterogeneous real-world networks by combining QKD and PQC under centralized management, ensuring tailored security for all nodes. The framework also applies communication network concepts to operate in an efficient, scalable, and resilient form. The framework leverages communication network principles to operate efficiently, scalably, and resiliently. It also extends the vKMS + QuSeC architecture to orchestrate security policies, assignments, and interoperability across QKD and PQC domains.
    \item \textbf{Adaptive Security Levels:} Definition of four security levels that integrate QKD and PQC flexibly, enabling quantum-safe protection in heterogeneous networks.
    \item \textbf{Standard-aware:} A standard-aligned model compatible with ETSI specifications, supporting incremental integration of QKD technologies into existing infrastructures without disrupting operation.
\end{enumerate}

The rest of the paper is organized as follows: Section \ref{sec:relatedwork} reviews prior work on QKD and PQC integration, and on extending quantum-safe security to heterogeneous networks. Then, Section \ref{sections/previousWork} summarizes our previous work on E2E quantum key establishment in QKD networks, on which this proposal builds. Subsequently, Section \ref{sec:proposed} introduces QuLore, the adaptive security framework, describing the system architecture and its operation. Section~\ref{sec:security_validation} summarizes the security analysis of the framework, and Section \ref{sec:validation} presents the implementation, testbed setup, and performance results. Finally, Section \ref{sec:conclusions} presents the main conclusions of the work.

\section{Related work}
\label{sec:relatedwork}

The combination of QKD and PQC for generating hybrid secure keys has attracted significant attention in recent years. The motivation of the hybrid approach is twofold: extending the range and applicability of QKD by complementing it with PQC mechanisms, and leveraging hybridization to obtain robust security guarantees, where the final session key remains secure as long as at least one of the contributing components is uncompromised.

Some theoretical foundations of hybrid Key Encapsulation Mechanisms (KEMs) are formalized in \cite{bindel2019hybrid}, where the security of several combiners, such as XOR-then-MAC (xTM) \cite{giacon2018kem}, dual-PRF \cite{bellare2024symmetric}, and the N Nested Dual-PRF, is introduced and proved. Their results can be used as the foundation for later works that integrate multiple key sources into secure session keys.

Building on this work, \cite{ricci2024hybrid} proposes a three-key combiner that takes as input 256-bit keys from different sources: a classical key generated via Diffie-Hellman, a PQC key generated with KYBER, and a QKD key. They implement and validate the proposal in an FPGA platform, and they also prove IND-CCA security. Similarly,  \cite{aquina2024quantum} introduces a quantum secure communication system leveraging a PRF-then-XOR split-key combiner, showing that security can be guaranteed even if only one of the underlying primitives remains secure.

Several works have demonstrated the feasibility of such combiners in realistic cryptosystems. For instance, \cite{zeng2024practical} defines a protocol for symmetric key establishment in QKD–PQC networks between two nodes, considering that several key distribution channels are available between them. The final key is derived through a hybrid KDF applied to multiple distribution channels. The obtained results validate the hybrid protocol where QKD and PQC interoperate within a joint quantum-classical network. In addition, \cite{garms2024experimental} presents Muckle++, a modular cryptosystem integrating hybrid authentication and hybrid key exchange. The proposed protocol derives the secure keys from combining QKD, PQC, and currently used public-key cryptography using a HMAC-based KDF (HKDF). The protocol is validated through real QKD hardware, FPGA-based PQC-KEM acceleration, and server-side implementations. Along similar lines, \cite{doring2022post} integrates PQC key exchange mechanisms into a QKD platform, enabling interoperability between the two technologies and providing a means to deliver quantum-safe keys to nodes where QKD is not available. They demonstrate the feasibility of the proposed scheme in the Berlin OpenQKD testbed, enabling quantum-safe communication between network nodes and extending the solution to wireless, last-mile, and end-user applications. These contributions highlight that hybrid protocols are both practical and deployable, bridging the gap between theoretical combiners and operational systems.

The role of KMSs in hybrid architectures has also been explored in detail. For instance, our previous work~\cite{sanz2025toward} proposed and validated a key management framework for QKD networks leveraging virtual KMSs under centralized control. Since that work provides the foundation for the approach presented here, it is explained in detail in Section~\ref{sections/previousWork}. Additionally, \cite{james2023key} provides a comprehensive analysis of KMS requirements for large-scale QKD networks, classical and quantum KMSs, and emphasizes the need for E2E guarantees even in the presence of compromised trusted nodes. In addition, to strengthen the overall security guarantees provided by a QKD network, they leverage PQC-based relaying of QKD keys between end nodes, with the final session key derived via hybridization at the KMS level. In a complementary direction, \cite{brauer2024linking}  demonstrates several approaches for achieving quantum-safe key exchange between three QKD testbeds in Europe, Berlin, Madrid, and Poznan, leveraging PQC-based encryption to emulate long-range QKD links. Different key exchange methods were defined to enable E2E communication between all the nodes of all the participating networks. In addition, they define a key hybridization process by applying a KDF as an internal process on the KMS that manages different internal key stores. These works illustrate how KMSs can serve as the central point for orchestrating hybridization in large-scale deployments. These efforts primarily focus on extending or strengthening fully QKD-enabled networks and providing a solid foundation for the combination of several input keys through KDFs. 

Additionally, some works also address environments where only a subset of nodes are QKD-capable. In this context, \cite{zhu2023qkd} introduces the concept of partially-QKD-deployed optical networks (pQKD-ON), where only part of the optical nodes are deployed with QKD systems, and defines the problem of key provisioning for secure E2E services in such networks. This work highlights the core challenge of enabling secure E2E communication in heterogeneous settings. They propose a multi-level key pool slicing scheme where QKD key pools are sliced, and some portions of keys are pushed outward via lightweight secure transfer. This approach enables having pre-shared QKD key-pairs between non-QKD-deployed nodes. Although they address key provisioning in heterogeneous networks, their approach lacks the centralized management and dynamic security level adaptation that our framework provides. Their solution focuses on static key pool slicing without the unified control plane and or standards compliance that enables real-world deployment. 

In summary, prior work has provided a strong foundation on secure key combiners for hybridization, practical hybrid protocols validated with QKD and PQC, and KMS architectures that support multi-domain integration. However, these efforts largely target homogeneous or fully QKD-enabled networks, not addressing the broader problem of dynamically extending quantum-safe security to heterogeneous environments where some nodes lack QKD capability. To the best of our knowledge, no existing framework provides a unified mechanism for adapting the security level depending on the node capabilities in heterogeneous and real-world networks. Our proposed framework addresses this gap by combining QKD and PQC under centralized management to deliver tailored, quantum-safe protection for all nodes, regardless of their underlying capabilities.

\section{Background: E2E quantum-safe key establishment in QKD networks}
\label{sections/previousWork}

This work builds on our previous work on scalable QKD architectures, which presented a foundational hierarchical E2E key establishment framework based on a centralized, programmable architecture leveraging Software-Defined Networking (SDN) principles. This work addressed the challenges of KMS identification, relay path discovery, and scalability in QKD networks. The proposed architecture introduced two key components that form the basis for the current work: the Virtual Key Management System (vKMS) and the Quantum Security Controller (QuSeC). 

On the one hand, the vKMS operates as an abstraction layer within each QKD-enabled node, managing multiple underlying KMS instances associated with different QKD modules. The vKMS presents a unified interface to applications, hiding the complexity of the QKD infrastructure and enabling efficient key retrieval and relay coordination. This abstraction enables applications to request quantum keys without needing knowledge of the underlying topology or KMS configuration.

On the other hand, the QuSeC operates as a centralized management entity that maintains a global view of the QKD network topology. Leveraging SDN principles, the QuSeC computes optimal E2E paths for key distribution, considering different metrics such as available key rate, hop count, or physical distance. It coordinates the installation of relay rules across multiple KMS instances and manages the complete lifecycle of multi-hop key relay operations. 

This hierarchical design supports both direct key delivery within direct QKD endpoints and multi-hop trusted relay across multiple nodes. The framework ensures compliance with ETSI specifications, including ETSI GS QKD 014 \cite{ETSI_GS_QKD_014} for key delivery, ETSI GS QKD 015 \cite{ETSI_GS_QKD_015} for SDN integration, and the ETSI GS QKD 020 \cite{ETSI_GS_QKD_020}, which is under development, for KMS interoperability in relay procedures. 

While this work improved scalability and provided a standard-aware key delivery mechanism for QKD networks, it assumed universal QKD capability across all network nodes, limiting its applicability to purely QKD-enabled environments. The current work addresses this limitation by extending the vKMS and QuSeC architecture beyond QKD-only deployments, enabling adaptive security management across heterogeneous networks where some nodes lack QKD infrastructure.

\section{QuLore: a framework for quantum-safe key establishment}
\label{sec:proposed}

This section presents \textbf{QuLore\footnote{QuLore is inspired by \textit{eguzkilore}, a traditional Basque flower that is hung on house doors as a protective symbol. In Basque mythology, the \textit{eguzkilore} represents the light of the sun, guarding the house and keeping witches and evil spirits away from home.}}: an adaptive framework that extends quantum-safe protection to all nodes within a network. The framework employs dynamic security assignment, adapting the level of security to the specific characteristics and capabilities of each node. In this context, the centralized network and security management plays a crucial role, providing a comprehensive view of the network to enable efficient and robust security management. 

The framework seamlessly integrates QKD and PQC to enable quantum-safe E2E key delivery across heterogeneous network environments. In addition, it includes a secure and efficient key provisioning mechanism allowing nodes without direct access to QKD links to participate in quantum‑safe sessions via relay and hybridization strategies. As a result, communications maintain resistance to quantum adversaries even when only a subset of the network is QKD‑enabled.

\subsection{\textbf{System Architecture}}

\begin{figure*}[h!]
\centering
    \includegraphics[width=1\textwidth]{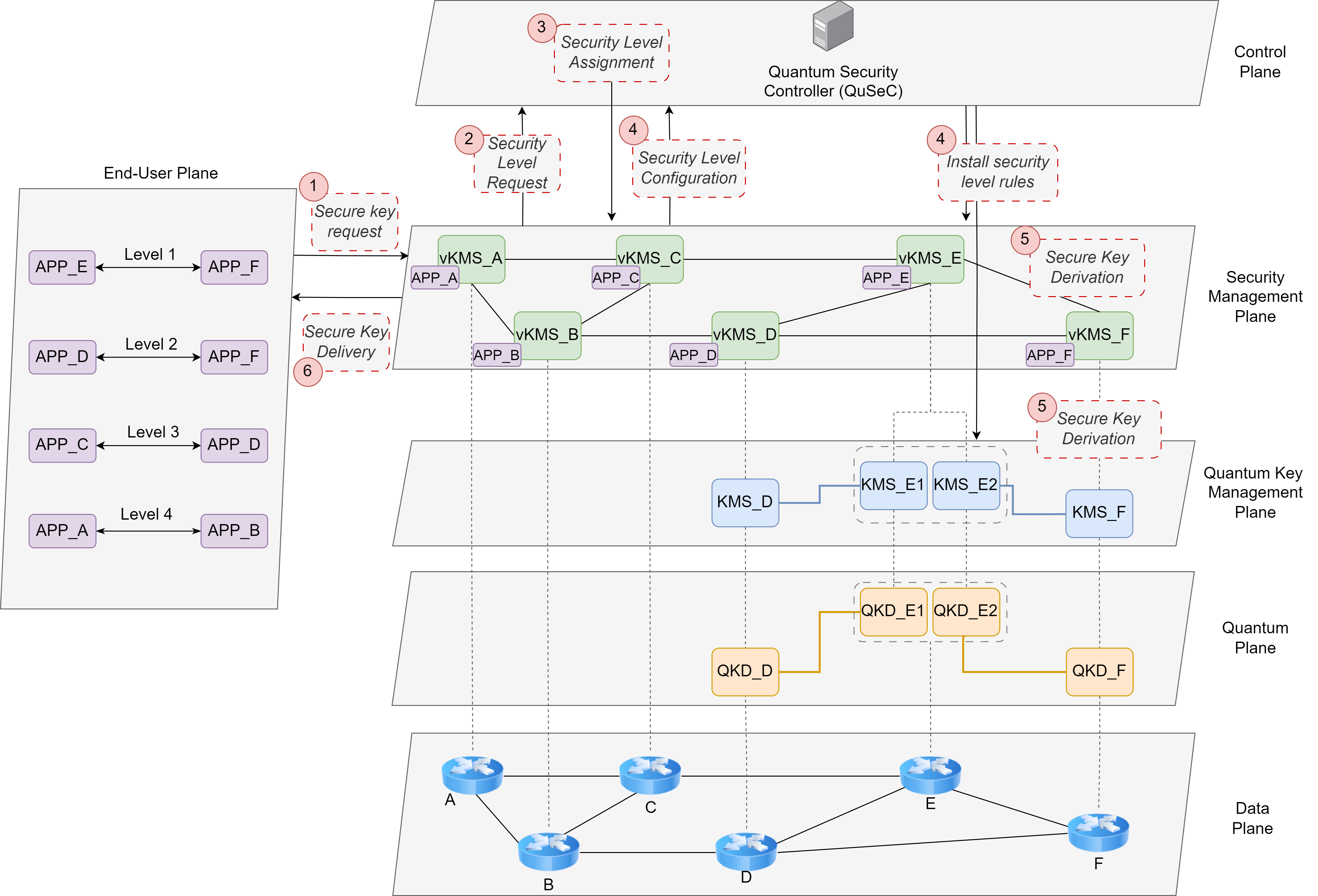}
\caption{System Architecture of the proposed framework.}
\label{fig:architecture}
\end{figure*}

The proposed architecture comprises two different node types. On the one hand, \textbf{QKD-enabled Nodes (QN)} are defined as nodes with access to QKD infrastructure capable of generating and exchanging QKD keys with peer QNs. Each QN hosts at least one KMS responsible for secure storage, synchronization, and delivery of QKD keys on demand. In Figure~\ref{fig:architecture}, nodes \textit{D}, \textit{E}, and \textit{F} are categorized as QN.

On the other hand, \textbf{Classical Nodes (CN)} lack QKD infrastructure, but are capable of employing PQC algorithms to achieve secure communication. In the example architecture shown in Figure~\ref{fig:architecture}, nodes \textit{A}, \textit{B}, and \textit{C} represent the CN category. 

Considering this, the proposed architecture is structured into several communication planes, as depicted in Figure~\ref{fig:architecture}:

\begin{itemize}
    \item \textbf{Data Plane:} This plane comprises the network devices, such as the switches or routers, used for the user traffic transmission.
    \item \textbf{Quantum Plane:} This plane comprises the physical QKD modules, including transmitters, receivers, and quantum links. In this plane, QKD keys are generated point-to-point between peer nodes and pushed to their associated KMSs.
    \item \textbf{Quantum Key Management Plane:} This plane comprises the KMSs responsible for managing QKD keys. Each QKD module in the Quantum Plane is paired with a KMS. The main functions of the KMS include secure storage and management of QKD keys, on-demand delivery of QKD keys to authorized applications, and synchronization with the peer KMS.
    \item \textbf{Security Management Plane:} A vKMS at every node, whether QN or CN, abstracts cryptographic operations from applications and implements the mechanisms instructed by the controller to derive a quantum-safe symmetric key between communication endpoints. The vKMS extends the functionalities of traditional KMSs by supporting key relay through quantum-safe channels, secure key generation via KDF, or PQC-based key establishment.    
    \item \textbf{Control Plane:} This plane comprises the QuSeC, which is responsible for managing both the network and security operations. In addition to standard SDN controller functions such as path computation or resource allocation, the proposed system introduces a \textbf{Security Level Assignment Function}. This function dynamically assigns and adjusts the security mechanism for the requesting communications based on the node types of the requesting nodes.
    \item \textbf{End-User Plane:} This plane includes the end-user applications that obtain quantum‑safe keys from the local vKMS for consumption by upper‑layer protocols.
\end{itemize}

This decomposition separates concerns: cryptographic material is never exposed to the QuSeC, applications are isolated from vendor‑specific KMS details, and policies are enforced uniformly across heterogeneous nodes.

\subsection{\textbf{Operational description of the security framework}}
The adaptive security framework ensures quantum-safe E2E communication by dynamically assigning Security Levels based on the involved nodes. The QuLore framework is designed to always provide the highest Security Level achievable, given the network capabilities and node types involved in the communication. 

\begin{figure*}[h!]
\centering
    \includegraphics[width=0.75\textwidth]{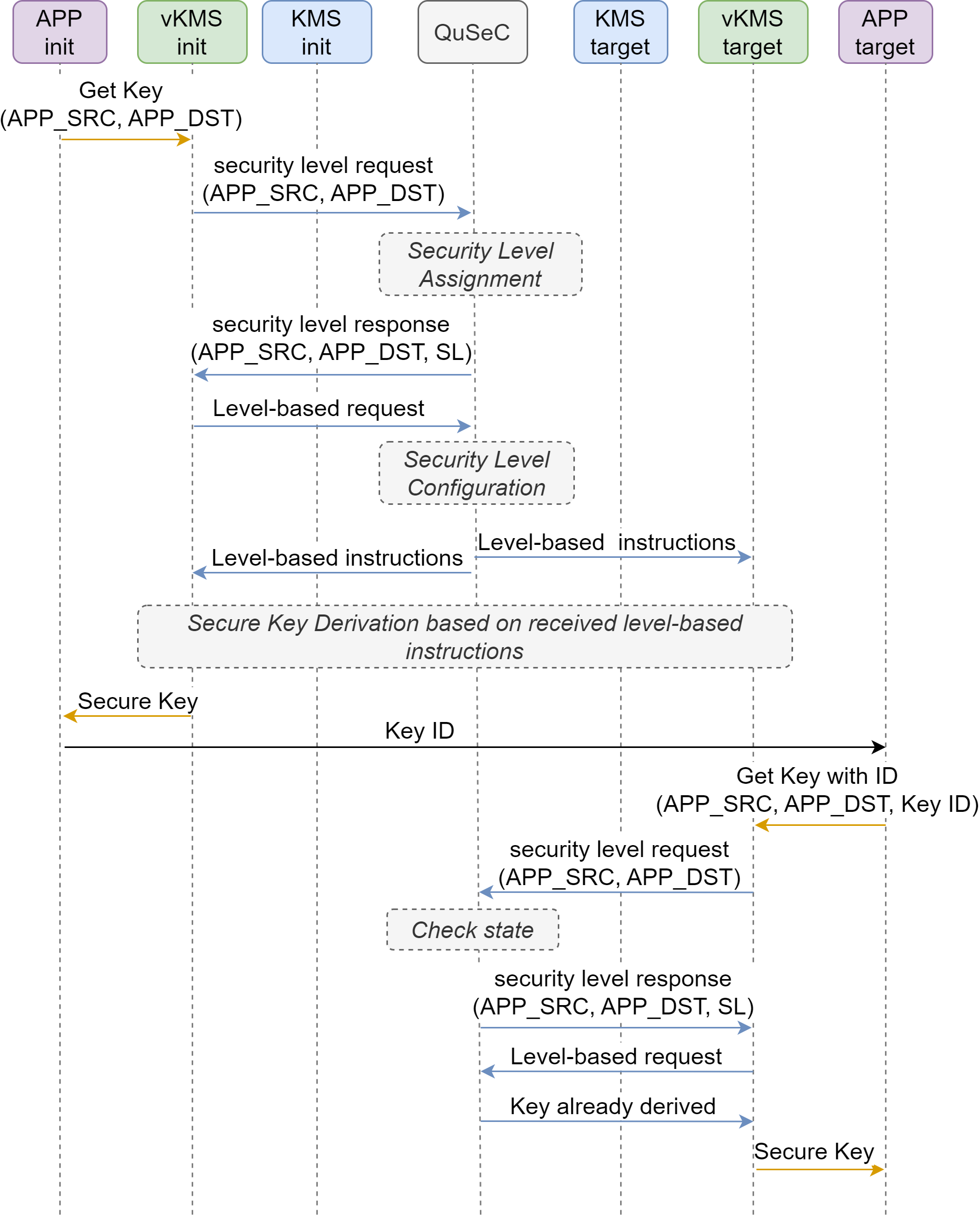}
\caption{Operational flow of the proposed framework for all levels.}
\label{fig:operationalFlow}
\end{figure*}

The complete E2E process operates through six sequential steps that transform an application's key request into a secure, quantum-safe session key. Figure~\ref{fig:operationalFlow} depicts the high-level operational flow of the proposed framework, illustrating how these steps interact across the different architectural planes. The process begins when an application initiates a secure session request and concludes with the delivery of quantum-safe keys to both communicating endpoints. 

\begin{enumerate}
    \item \textbf{Secure Session Request.} An initiating application requests a quantum-safe communication session with another application to its associated vKMS. This request is made by sending a \texttt{Get Key} message defined in the ETSI GS QKD 014 API. The vKMS receives this request and prepares to coordinate with the QuSeC to deliver the secure key.
    
    \item \textbf{Security Level Request.} The source vKMS cannot independently determine the optimal security approach since this depends on the capabilities of both endpoints and the quantum path availability. Therefore, upon receiving the request, the vKMS queries the QuSeC by sending a \texttt{security\_level\_request (APP\_SRC | APP\_DST)} message to know the Security Level that corresponds to the communication. 
    
    \item \textbf{Security Level Assignment.} The QuSeC executes the \textbf{Security Level Assignment Function} to determine the best security level. It resolves endpoint capabilities (QN or CN) and path feasibility, selects one of the four Security Levels (detailed in Section \ref{sec:assignment}) and mechanisms, and responds to the vKMS with the assigned level.
    
    \item \textbf{Security Level Configuration.} With the Security Level determined, the vKMS requests detailed configuration instructions from the QuSeC. The QuSeC then computes a comprehensive configuration policy with instructions tailored to the selected Security Level. These instructions range from minimal relay path installation to PQC or KDF configuration. The QuSeC distributes these configuration policies to all involved vKMSs and KMSs, programming the distributed system to execute the chosen security mechanisms.
    
    \item \textbf{Secure Key Derivation.} This is the execution phase where all configured components work together to generate the shared session key. The specific operations depend on the assigned Security Level: purely QKD-based operations may involve retrieving quantum keys from their KMSs, relay operations may forward keys through intermediate trusted nodes using OTP encryption, PQC operations may establish shared secrets using post-quantum KEMs, and hybrid approaches may combine multiple key sources through cryptographic KDFs. Throughout this process, each vKMS performs its assigned role while maintaining synchronization with its peer to ensure both endpoints derive the identical session key.
        
    \item \textbf{Secure Key Delivery.} Once the quantum-safe session key has been successfully derived, each vKMS delivers the key to its respective application over a secure, authenticated channel. The applications can then proceed with their secure communication using the established session key. 
\end{enumerate}

The key identifier exchange between applications occurs after step 6, when the initiating application transmits the session key identifier to the target application through their application-level protocol. When the target application subsequently requests its copy of the session key using a \texttt{Get Key with ID} message, the process is streamlined: the QuSeC maintains session state and recognizes this as a session key retrieval rather than a new session establishment, and instructs the target's vKMS to deliver the previously derived key directly, bypassing steps 4 and 5. The mechanism by which the key identifier is conveyed is out of the scope of this work.

After both applications have received the secure keys, they can proceed with communication using the established session key. The QuLore framework is designed to be protocol-agnostic, allowing the quantum-safe key to be integrated into different security mechanisms, such as pre-shared keys, session keys, or encryption keys, depending on the requirements of the underlying communication protocol.

The following subsections describe in more detail the specific procedures of Security Level Assignment, Security Level Configuration, and Secure Key Derivation for each of the four defined Security Levels.

\subsubsection{Security Level Assignment.}
\label{sec:assignment}
\begin{figure*}[t]
  \centering
  \subfloat[Security Level 1]{%
    \includegraphics[width=0.44\textwidth]{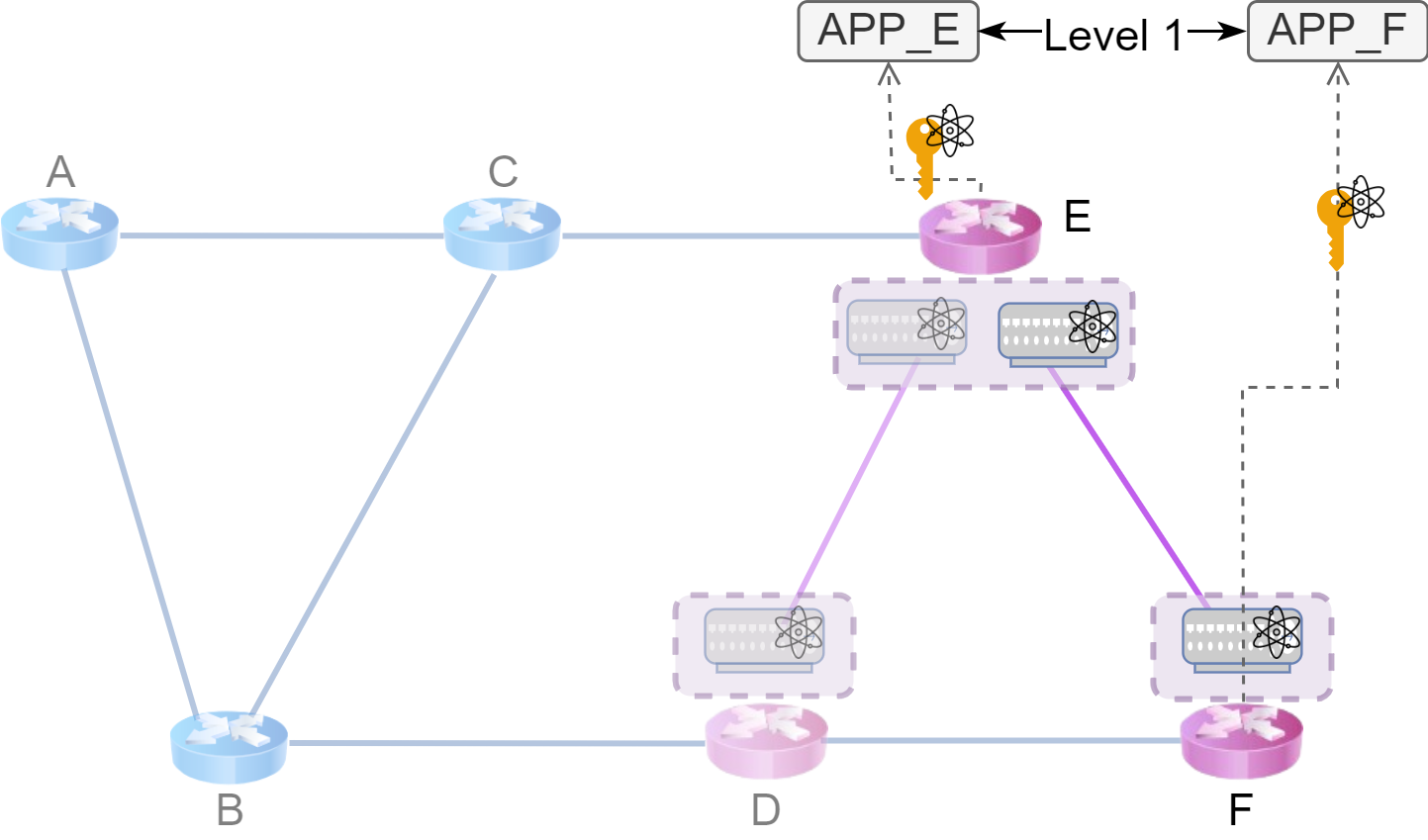}%
    \label{fig:l1}
  }\hfill
  \subfloat[Security Level 2]{%
    \includegraphics[width=0.44\textwidth]{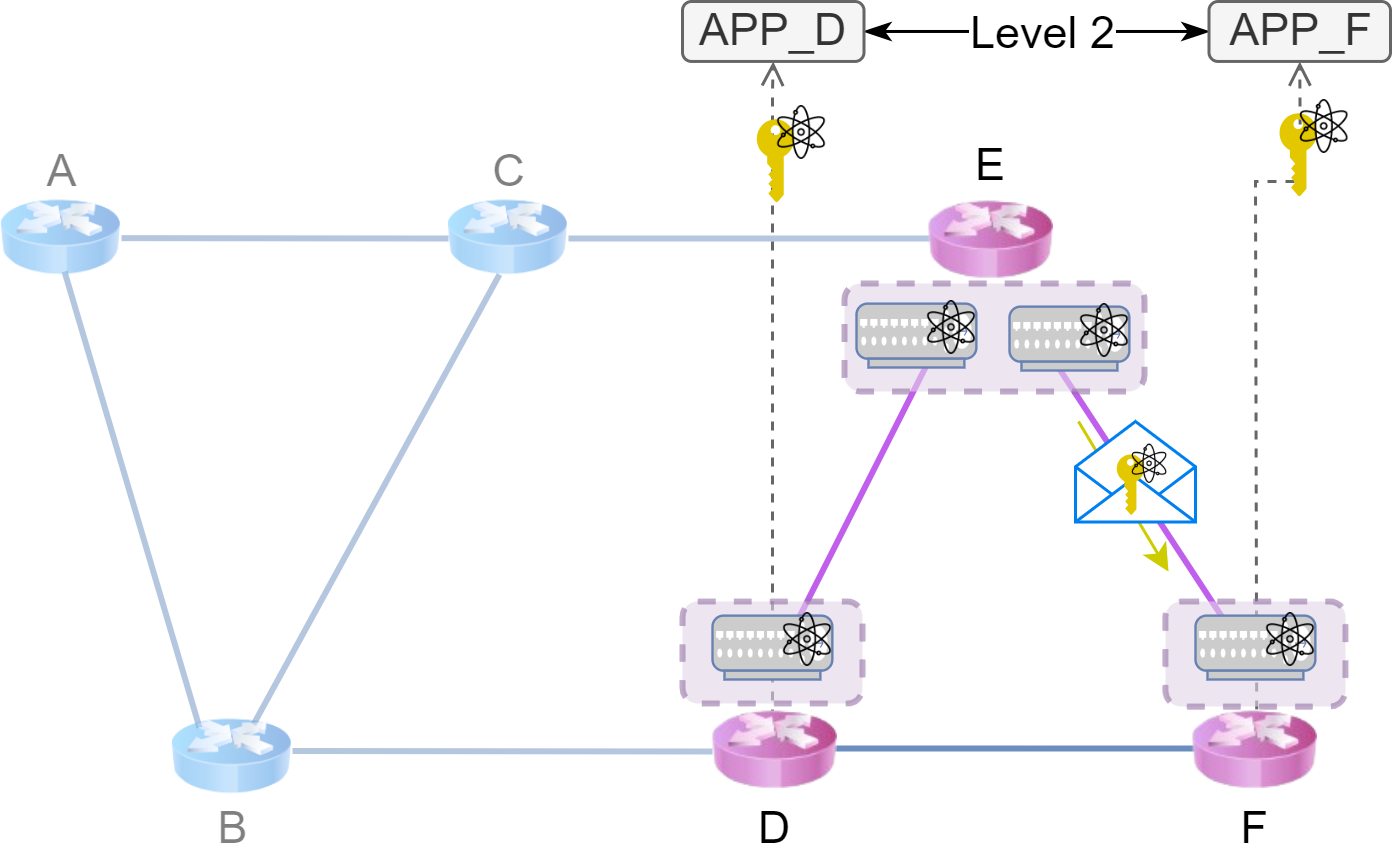}%
    \label{fig:l2}
  }\\[0.5em]
  \subfloat[Security Level 3]{%
    \includegraphics[width=0.44\textwidth]{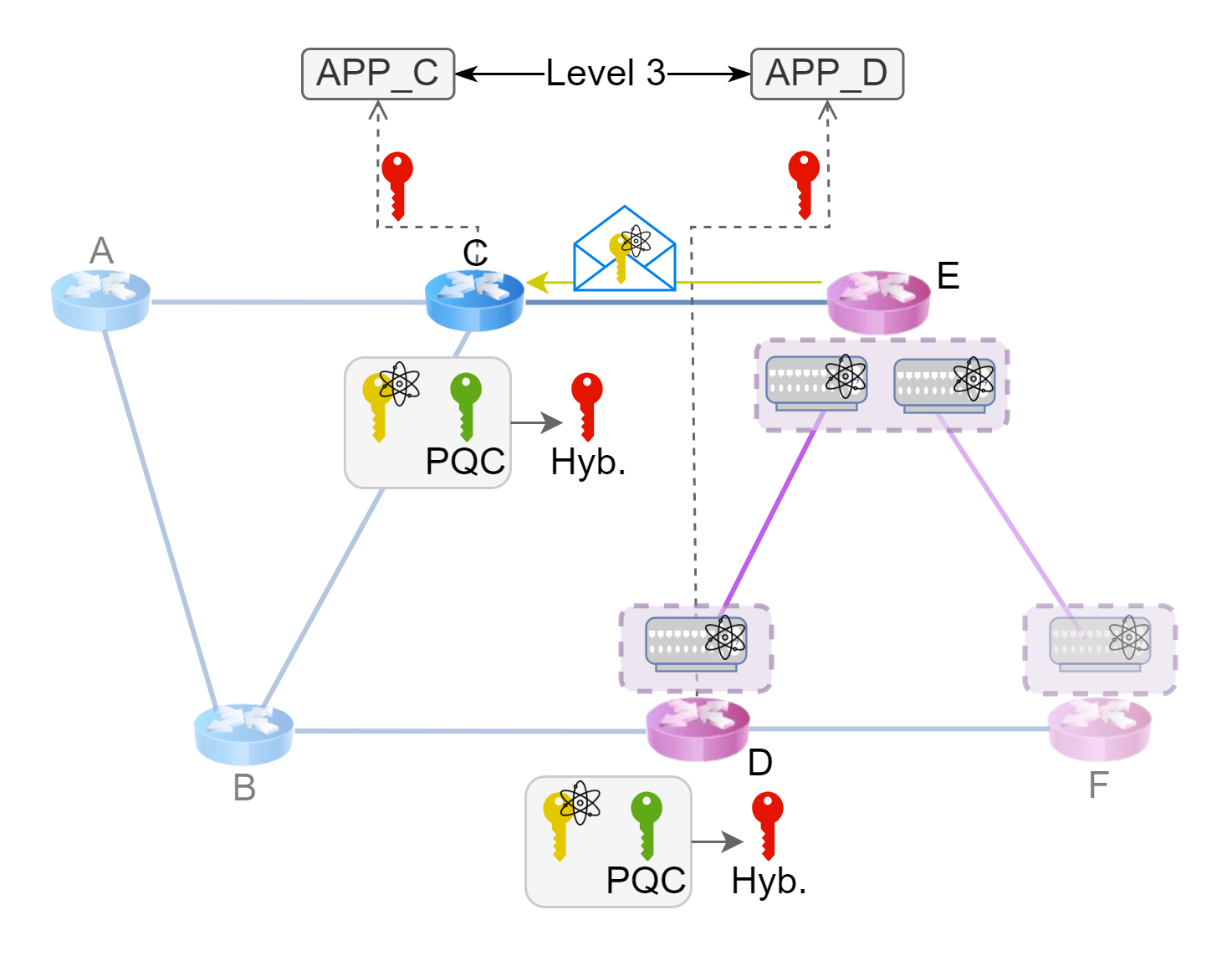}%
    \label{fig:l3}
  }\hfill
  \subfloat[Security Level 4]{%
    \includegraphics[width=0.44\textwidth]{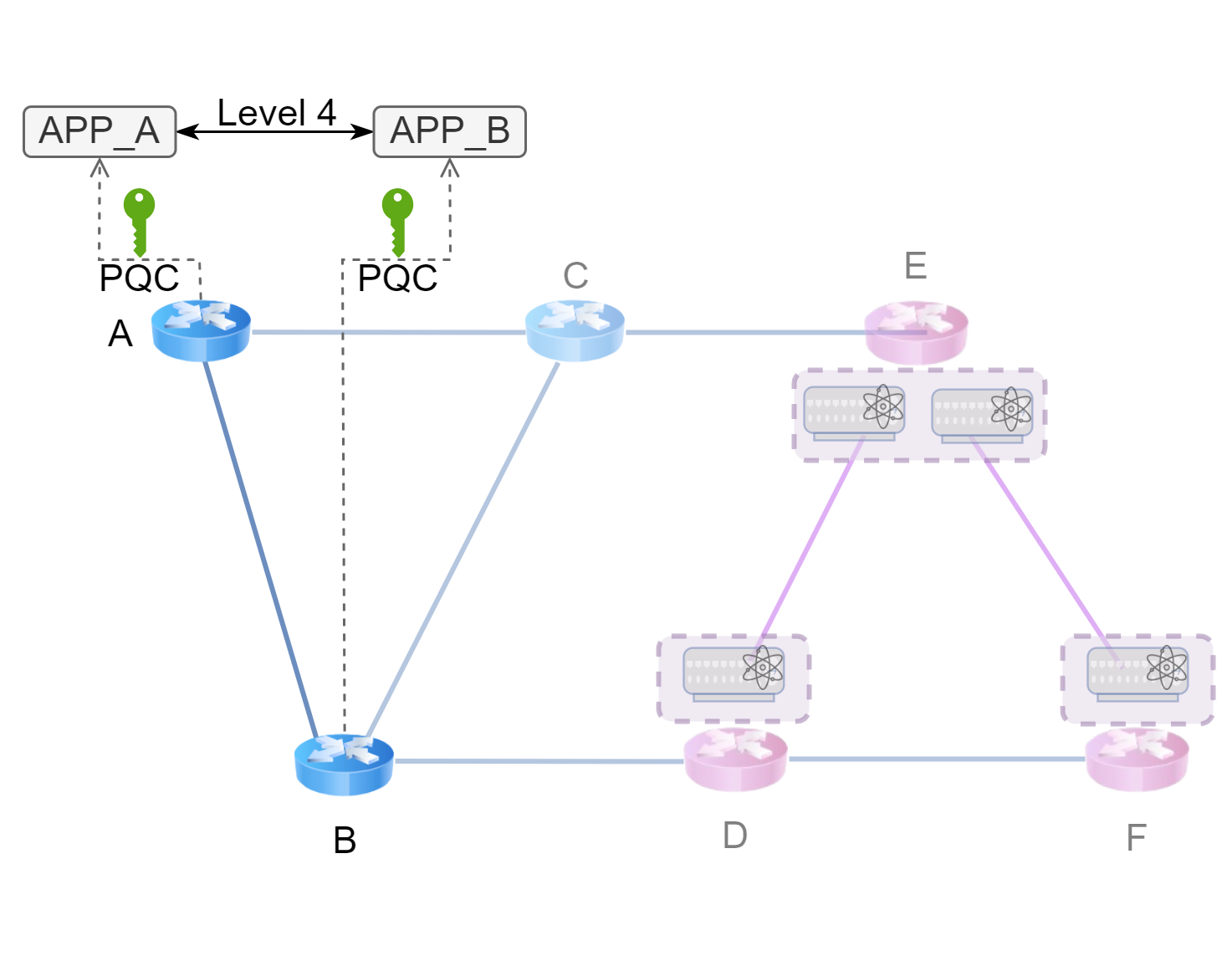}%
    \label{fig:l4}
  }

  \vspace{1ex}

\includegraphics[width=0.35
\textwidth]{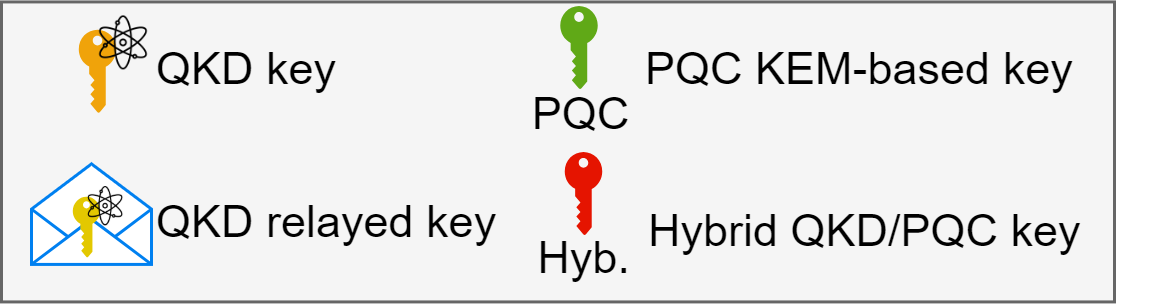}

  \caption{High-level operation of each Security Level.}
  \label{fig:four-with-legend}
\end{figure*}

Each vKMS, upon receiving a secure key request from an application, queries the QuSeC to request a Security Level for the communication between the initiating and the target applications. The proposed framework adapts the Security Level based on whether the involved applications belong to QN or CN, and on the quantum path availability. The QuSeC assigns a Security Level to each E2E communication and coordinates all the associated vKMSs and KMSs to execute the required operations to derive a symmetric E2E key with the highest security possible. In this context, after analyzing all possible scenarios, four different Security Levels have been defined:

\begin{itemize}
    \item \textbf{Security Level 1 - Direct QN-to-QN.} This Security Level applies when two applications, each belonging to QNs directly connected via a QKD link, request secure communication. The assigned Security Level is ITS, as no classical key exchange is involved, thereby fully preserving the quantum-safe properties of QKD. Figure~\ref{fig:l1} shows the high-level operation of Level 1, between applications on nodes E and F.
    \item \textbf{Security Level 2 - Multi-hop QN-to-QN.} This Security Level is assigned when two applications on QNs without a direct QKD link require secure communication. In such cases, the QKD with a trusted relay approach is employed, where keys are forwarded through intermediate trusted nodes leveraging One-Time Pad (OTP) encryption. Within the declared trust boundary for relays and with secure control and API channels, secrecy arguments are information-theoretic for the key material. Figure~\ref{fig:l2} illustrates this Security Level for applications on nodes D and F.
    \item \textbf{Security Level 3 - CN-to-QN.} This level applies when one application resides on a node without QKD infrastructure (CN). In this case, a relay and hybridization approach is employed to ensure secure key generation. Specifically, a hardened KDF derives a session key from at least one QKD-derived segment and one PQC KEM shared secret. The resulting key remains secure as long as at least one input key is secure. This multi-source approach produces a quantum-resistant symmetric key by combining independent security mechanisms. Figure~\ref{fig:l3} depicts the high-level operation of Level~3 between applications on nodes C and D, including the relay of a QKD key via node E and the hybridization process.
    \item \textbf{Security Level 4 - CN-to-CN.} This Security Level applies when both communicating applications reside on CNs, meaning no QKD infrastructure is available. In this case, a purely PQC-based approach is employed to ensure secure key exchange and derivation. Additionally, a KDF hybridization approach can be leveraged by combining multiple cryptographic sources to further enhance resilience. While this level does not provide ITS as in QKD-based approaches, it ensures robust protection against both classical and quantum threats. Figure~\ref{fig:l4} shows this Security Level for applications on nodes A and B.
\end{itemize} 

In practice, selection follows a simple preference ordering: $Level\ 1 > Level\ 2 > Level\ 3 > Level\ 4$, subject to quantum path feasibility and node capacity. 

\subsubsection{Security Level Configuration.}
\label{sec:configuration}
After level assignment, the vKMS sends a level-based request to the QuSeC, which computes a level-specific configuration that encodes the session context, the selected cryptographic suites, and the minimal set of instructions required by each participant. For Levels 1 and 2, this step has been described in detail in our previous work~\cite{sanz2025toward}, and it primarily involves KMS identification and, when applicable, relay path installation. For Levels 3 and 4, however, additional structure is required.

In the case of Level 3, the configuration is built around three roles that reflect the functions of the participating vKMSs. The \textbf{receiver} role is for the vKMS at the classical endpoint, it is the only endpoint that will receive, decrypt, and unveil the relayed QKD key. It will also establish a PQC KEM-based key with the passive vKMS, and it will derive and deliver the final session key to its application. The \textbf{passive} role is for the vKMS at the QKD-enabled endpoint, and it is responsible for retrieving the QKD key from its KMS, establishing a PQC KEM-based secret with the receiver, and deriving and providing the session key to its own application. Between them, the \textbf{relay} role is for the intermediate KMS/vKMS pair at the QN that has a QKD link with the passive node. It performs the relay of the QKD key via PQC KEM encapsulation and OTP encryption to the receiver. 

Although all participants receive a session configuration message with the same overall structure, the content of the message is not identical for every participant. Each message includes common session information, such as a unique session identifier, as well as a policy block tailored to the role of the recipient. This policy block specifies the actions to be performed, the selected cryptographic suites, and the identities of the relevant endpoints.

The configuration for Level 4 is comparatively simpler. Here, both endpoints directly establish a shared secret using the configured PQC KEM. The QuSeC provides each endpoint with a bilateral set of instructions that specify the KEM suites to be used, and, when multiple input keys are present, the KDF recipe for combining them into the final session key.

\subsubsection{Secure Key Derivation.}
\label{sec:derivation}
Once the configuration is installed, Secure Key Derivation proceeds directly according to the installed policy instructions. As previously mentioned, the procedures corresponding to Levels 1 and 2 have been introduced in the previous work. In summary, Level 1 consists of retrieving a QKD key from the KMS. Level 2 also involves retrieving a QKD key from the KMS, but adds a key relay procedure between different KMS instances, following the relay path configured by the QuSeC in the preceding step. 

\begin{figure*}[h!]
\centering
    \includegraphics[width=0.7\textwidth]{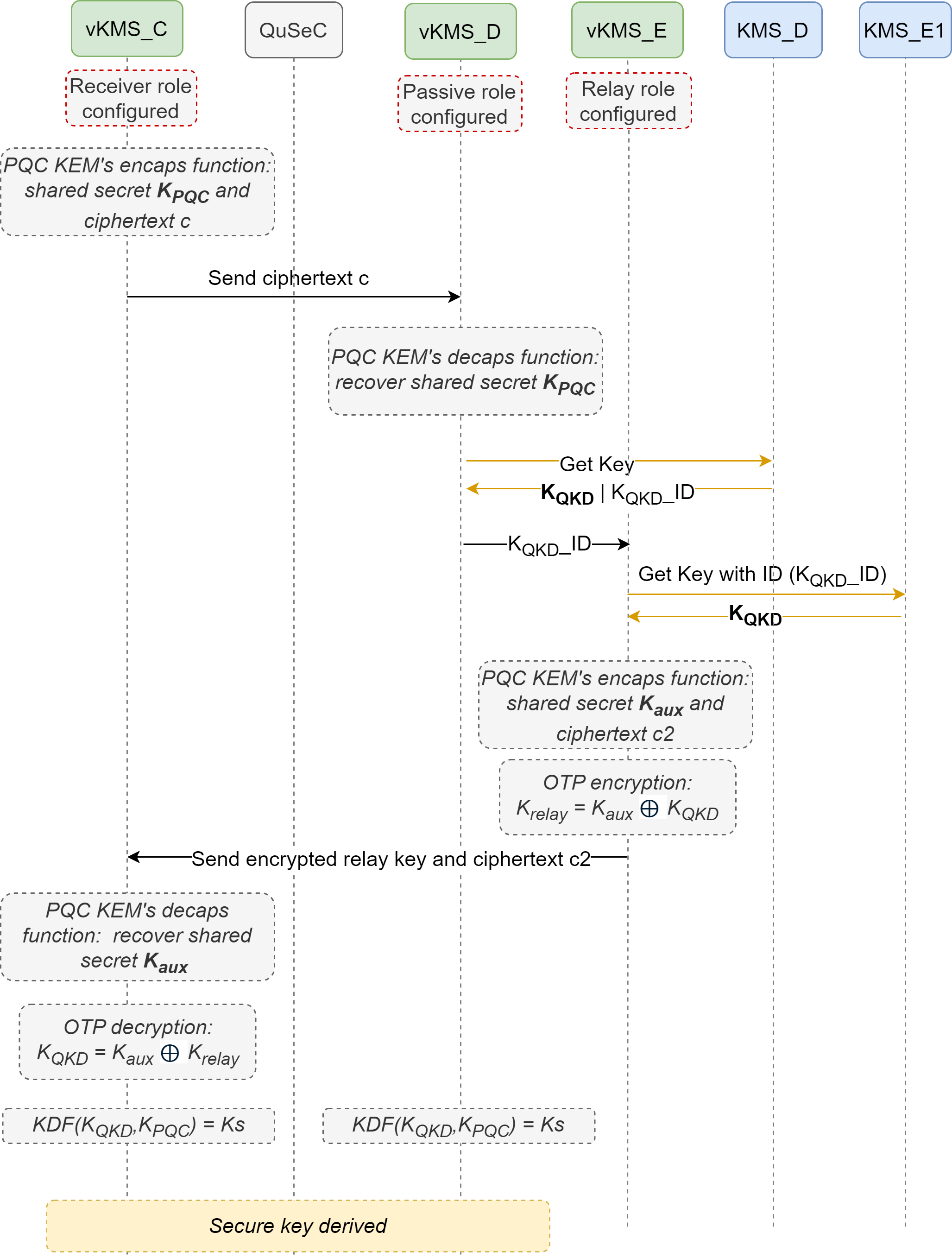}
\caption{Operational flow of Secure Key Derivation in Level 3.}
\label{fig:level3-keyDerivation}
\end{figure*}

For Level 3, the role-scoped steps that must be performed are described as follows, and they are graphically represented in Figure~\ref{fig:level3-keyDerivation}:

\begin{enumerate}
    \item The receiver and the passive vKMS establish a shared secret by running the configured PQC KEM. In this process, the passive generates a public-private key pair and shares the public key with the receiver vKMS. The receiver vKMS then performs the encapsulation function using the passive's public key to generate a shared secret and corresponding ciphertext. The ciphertext is then transmitted back to the passive vKMS, which performs the decapsulation function using its private key to recover the same shared secret, thereby establishing a PQC key between both endpoints.
    \item The passive vKMS retrieves a QKD key from the KMS through the ETSI GS QKD 014 key delivery API via a \texttt{Get key} message, and transmits the corresponding key ID to the relay vKMS.
    \item Upon receiving the key identifier, the relay vKMS retrieves the identical QKD key using the \texttt{Get key with ID} message of the ETSI GS QKD 014 API. 
    \item The relay vKMS transports the key using post-quantum cryptography: it performs the encapsulation function of the configured PQC KEM with the receiver's public key to generate an auxiliary shared secret with the receiver and a ciphertext. The relay then applies OTP encryption between the auxiliary key and the QKD key. It transmits both the ciphertext and the encrypted key to the receiver vKMS.    
    \item The receiver vKMS performs the decapsulation of the received ciphertext using its private key to recover the auxiliary shared secret. It then performs the OTP decryption to recover the original QKD key.
    \item Both the receiver and the passive vKMS apply the configured KDF using as input the KEM secret and the QKD key in order to derive the session key.
\end{enumerate}

\begin{figure}[t!]
\centering
    \includegraphics[width=0.9\columnwidth]{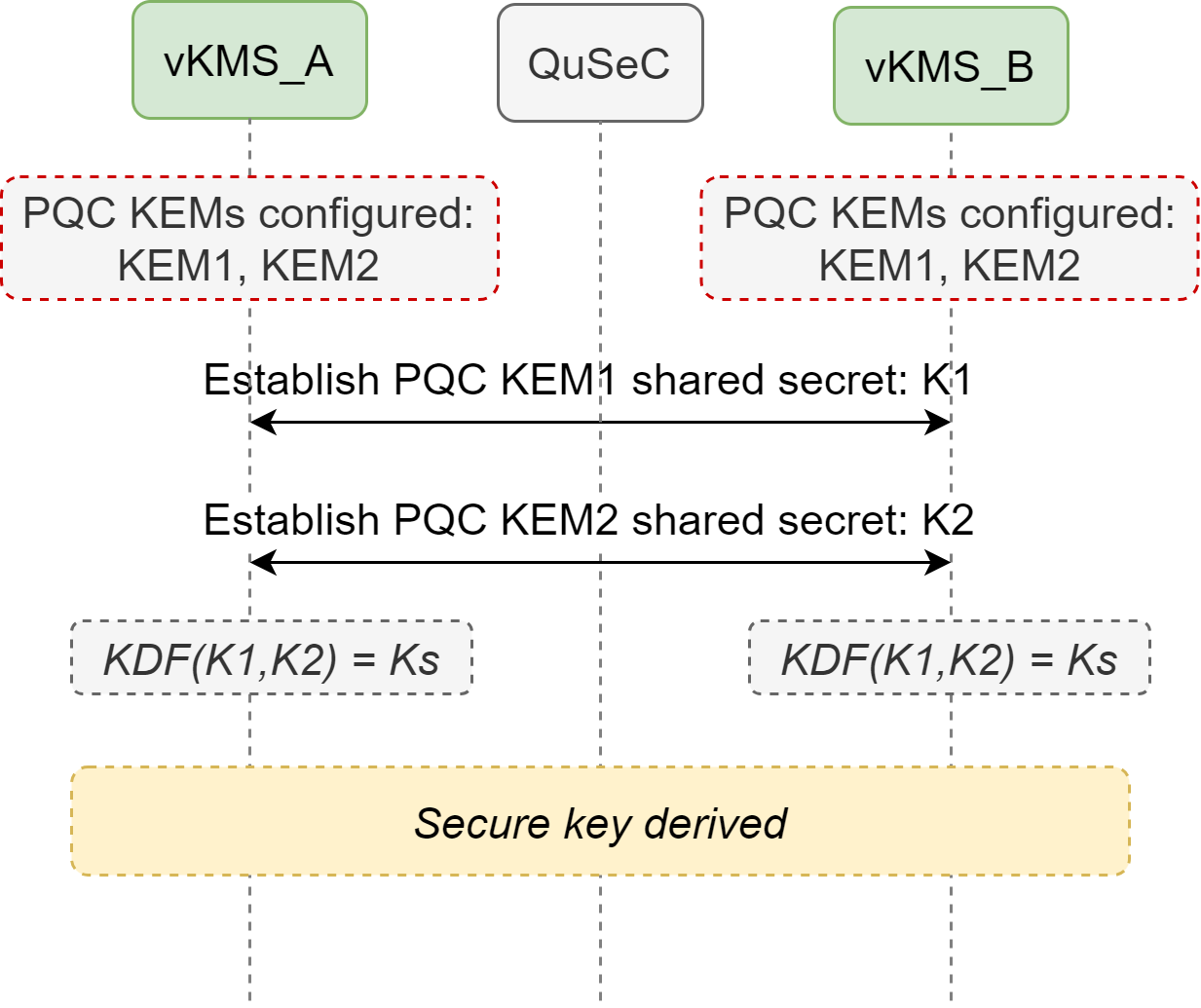}
\caption{Operational flow of Secure Key Derivation in Level 4.}
\label{fig:level4-keyDerivation}
\end{figure}

On the other hand, the configuration of Level 4 entails the following operations:

\begin{enumerate}
    \item Both vKMS endpoints execute the configured KEM to generate a shared secret.
    \item Optionally, they can incorporate dual KEMs or additional entropy sources if enabled by policy. In this case, all generated shared secrets are combined using the configured KDF to derive the final session key.
\end{enumerate}

Figure~\ref{fig:level4-keyDerivation} depicts the operational flow for the key derivation in level 4, for a configuration where the QuSeC configured a policy with dual KEM enforcement.




\section{Security analysis}
\label{sec:security_validation}
\newcommand{\Adv}{\mathsf{Adv}}
\newcommand{\INDSK}{\mathsf{IND\text{-}SK}}
\newcommand{\AdvIND}[1]{\mathsf{Adv}^{\INDSK}_{#1}}

This section presents an overview of the security analysis of the QuLore framework under a Dolev-Yao network model~\cite{dolev2003security}: the adversary can eavesdrop, delay, replay, inject, and arbitrarily modify any message transmitted over the classical network. Cryptographic primitives are analyzed in the standard computational model, considering a possibly quantum probabilistic-polynomial time (PPT) adversary, and capturing its ability to break the Secure Key Derivation procedure. The security analysis relies on the following security assumptions:

\begin{enumerate}
    \item Authenticated secure channels: All control/management and key-delivery communications between applications, vKMSs, KMSs, and the QuSeC are transmitted over mutually authenticated and confidential channels.

    \item ITS for QKD keys: Each QKD link generates information-theoretically secure random keys under standard QKD assumptions, including an authenticated classical channel for QKD post-processing.

    \item Security of PQC and KDF: PQC KEMs are Indistinguishable under Adaptive Chosen-Ciphertext Attacks (IND-CCA2) against classical/\linebreak[4]quantum adversaries. Similarly, the KDF is modeled as PRF-like on its inputs, generating an indistinguishable from uniform output as long as at least one input remains secret.

    \item No compromised nodes: The threat model excludes malicious insiders and compromised nodes. The attacker does not access internal states of entities, cannot extract keys from memory, and cannot perform physical extraction attacks. Trusted relay nodes in Levels 2–3 (Levels 1 and 4 do not comprise the use of trusted relays) are inside a declared trust boundary.

    \item Controller key isolation: Architecturally, session keys and intermediate key material are never transmitted to QuSeC, which only sees metadata, IDs, and policies. Even if QuSeC were compromised beyond the primary model, it cannot directly learn session keys from protocol messages.
    
\end{enumerate}

Under these assumptions, the security analysis focuses on session-key confidentiality, captured via an indistinguishability game (IND-SK): after interacting with honest protocol instances over the network, the adversary selects a completed test session and must distinguish the real session key from a uniformly random string of equal length. The protocol achieves confidentiality if the adversary’s advantage is negligible in the security parameter.

For each Security Level, full game-based proofs and reductions are performed, leading to the confidentiality guarantees summarized in Table~\ref{tab:confidentiality}. For completeness, short proof sketches and theorems are included in this section. 

In \textbf{Level 1}, the established key is directly a QKD-generated secret $K_{QKD}$. For an adversary to distinguish the real session key from random is by either eavesdropping on QKD channels (information-theoretically prevented by QKD), breaking the KMS protection (violating trust assumption), or recovering from network traffic (not possible due to authenticated and encrypted channels). Therefore, since the adversary's information about the key is bounded information-theoretically, no computational advantage allows the adversary to distinguish the real key from random data, hence:

\[ \AdvIND{\text{Level 1}}(\lambda) \leq \text{negl}(\lambda) \]

In \textbf{Level 2}, the established key is a QKD-derived secret $K_0$ that is relayed hop-by-hop using OTP encryption under independent per-link QKD keys. An adversary observing the classical network only sees relay ciphertexts $C_i = K_0 \oplus K_{QKD,link_i}$, which are uniformly random due to OTP when the pad is secret and fresh, so they reveal no information about $K_0$. Therefore, the only practical way for the adversary to distinguish the real E2E session key from random is to (i) eavesdrop on QKD keys (information-theoretically prevented by QKD), (ii) compromise a trusted relay or access internal KMS/vKMS memory (violates the trust-boundary assumption), or (iii) recover key material from network traffic (not possible because key transfers inside nodes use mutually authenticated confidential channels). Hence, the adversary’s information about the session key remains bounded information-theoretically, and its IND-SK advantage is negligible under the stated assumptions, leading to the following theorem proof:

\[\AdvIND{\text{Level 2}}(\lambda) \leq \text{negl}(\lambda)\]

In \textbf{Level 3}, the session key is derived as $K_{session}=KDF(K_{QKD} || K_{PQC})$, where $K_{PQC}$ comes from a PQC KEM exchange and $K_{QKD}$ is obtained from the QKD link and relayed to the CN using OTP under an auxiliary PQC-derived secret $K_{PQC,aux}$. From the adversary’s perspective, the only way to distinguish the real session key from random is to either (i) break the IND-CCA2 security of the KEM to recover $K_{PQC}$ (and potentially $K_{PQC,aux}$), (ii) distinguish the KDF output from random, breaking the PRF-like property of the KDF, (iii) violate the QKD secrecy assumptions or compromise internal KMS/vKMS state (out of scope), or (iv) recover secrets from protected network traffic (not possible under the secure-channel assumption). Therefore, any non-negligible adversary can be reduced to breaking IND-CCA2 (KEM) or PRF (KDF), so the IND-SK advantage is negligible under those security assumptions, as stated in the following theorem:

\[\AdvIND{\text{Level 3}}(\lambda) \leq \Adv^{\text{IND-CCA2}}_{\text{KEM}}(\lambda) + \Adv^{\text{PRF}}_{\text{KDF}}(\lambda) + \text{negl}(\lambda)\]

In \textbf{Level 4}, both endpoints are classical and establish the session key purely using PQC KEM, and optionally a KDF if combining multiple KEM outputs. For the single-KEM variant, an adversary can only distinguish the real session key from random by breaking the KEM’s IND-CCA2 security, or violating the protected-channel/trusted-node assumptions. Otherwise the KEM shared secret is computationally indistinguishable from uniform, and so is the derived session key. For the dual-KEM configuration, the adversary’s task is even harder: to distinguish the final key, it must break at least one of the assumed components: IND-CCA2 of KEM1, IND-CCA2 of KEM2, or PRF security of the KDF. Otherwise, the combined output remains indistinguishable from random. Therefore, under IND-CCA2 security of the employed KEM(s) and PRF-like KDF when used, the adversary’s distinguishing advantage in IND-SK is negligible. The following theorems show the advantage of an adversary for single KEM and dual KEM cases, respectively:

\[    
    \AdvIND{\text{Level 4 (single)}}(\lambda)
    \leq \Adv^{\text{IND-CCA2}}_{KEM}(\lambda)
       + \text{negl}(\lambda)
\]

\[
    \AdvIND{\text{Level 4 (dual)}}(\lambda) 
    \leq \Adv^{\text{IND-CCA2}}_{\text{KEM}_1} (\lambda)
    + \Adv^{\text{IND-CCA2}}_{\text{KEM}_2} (\lambda) \]
\[
    + \Adv^{\text{PRF}}_{\text{KDF}} (\lambda)
    + \text{negl}(\lambda)
\]

    
    
    
    
    
    
    

\begin{table*}[t]
\centering
{\scriptsize
\captionsetup{justification=justified,singlelinecheck=false}
\caption{Confidentiality guarantees per Security Level under the threat model.}
\label{tab:confidentiality}

\begin{minipage}{\textwidth}
\centering
\begingroup
\renewcommand{\arraystretch}{1.5}
\setlength{\tabcolsep}{6pt}
{
\begin{tabularx}{\linewidth}{
>{\centering\arraybackslash}m{0.06\linewidth} 
>{\raggedright\arraybackslash}X 
>{\raggedright\arraybackslash}X 
>{\raggedright\arraybackslash}X}
\hline
\textbf{Level} & \textbf{Construction} & \textbf{Confidentiality guarantee} & \textbf{Main assumptions} \\
\hline

L1 & Direct QKD key delivery
   & Information-theoretic secrecy
   & QKD assumptions; protected key delivery channels \\

L2 & Multi-hop relay using OTP on per-link QKD keys
   & Information-theoretic secrecy within trust boundary
   & Trusted relays (no compromise), fresh OTP keys per hop \\

L3 & Hybrid: $K = KDF(K_{QKD}\,\|\,K_{PQC})$ with QKD relay protected by PQC-derived auxiliary secret
   & Computational IND-SK confidentiality
   & IND-CCA2(KEM), PRF(KDF), trusted relay \\

L4 & PQC-only (single KEM or dual-KEM + KDF)
   & Computational IND-SK confidentiality
   & IND-CCA2(KEM), PRF(KDF) \\

\hline
\end{tabularx}
}
\endgroup


\end{minipage}
}
\end{table*}

\section{Performance evaluation}
\label{sec:validation}

This section presents the implementation used to validate and evaluate the performance of the QuLore framework. First, the testbed and implementation details are introduced. Next, the functional validation of the proposed security framework is presented, demonstrating its correct operation. Finally, performance evaluation results are discussed.

\subsection{\textbf{Implementation and Testbed Setup}}
All system components were fully implemented, including the vKMS, KMS, and QuSeC modules, following the architectural design presented in Section \ref{sec:proposed} and ensuring full compliance with ETSI GS QKD 014 specification.

The proposed solution was deployed on a Supermicro SYS-E200-8D server, featuring an Intel(R) Xeon(R) D-1528 processor running at 1.90 GHz on an x86\_64 architecture with 6 cores, backed by 64 GB of RAM and approximately 500 GB of local storage. On this infrastructure, an experimental testbed that replicates the heterogeneous network topology depicted in Figure~\ref{fig:architecture} was deployed using Ubuntu 20.04 and Kubernetes orchestration. The testbed comprises six nodes: three Classical Nodes (CN A, B, C) and three QKD-enabled Nodes (QN D, E, F). 

The most relevant implementation details for each component are as follows.

\begin{itemize}
    \item Each vKMS instance supports the complete ETSI GS QKD 014 API to interface with end applications, extended with mechanisms for communication with the QuSeC to receive Security Levels and the related instructions. They support adaptive security mechanisms, including PQC integration and hybrid key derivation, under policy-driven Security Level execution. Post-quantum operations were realized with standardized algorithms, specifically ML-KEM~\cite{FIPS203} for KEM operations.
    \item The KMS instances also implement the ETSI GS QKD 014 API for interaction with the vKMSs for QKD key delivery, as well as the ETSI GS QKD 020 API draft for KMS interoperability in multi-hop QKD scenarios. In addition, they support the operations proposed in our previous work for communication with the QuSeC and the vKMS.
    \item The QuSeC implements the Security Level Assignment Function and associated configuration tasks, including topology-awaree path computation for multi-hop QKD scenarios, and automated policy generation for each Security Level. The path computation assumes equal link costs, so the routing criterion corresponds to the minimum hop count, although these weights could be dynamically adjusted to reflect additional metrics such as available key rate.
    \item The QKD modules of the quantum plane have been implemented using SimulaQron, a distributed quantum network simulator designed for software development. However, as the framework relies on standardized ETSI APIs, it maintains full compatibility with real QKD hardware from any vendor. The simulator serves purely as a proof-of-concept platform, enabling comprehensive validation without requiring physical infrastructure. Performance validation in this work targets the proposed framework, which is independent of the underlying quantum-channel key rate. Therefore, SimulaQron is used solely as a key source to populate the KMSs, and can be transparently replaced by real QKD hardware without affecting the reported performance results.
\end{itemize}

Finally, to ensure consistency and reproducibility, all cryptographic operations are configured with standardized parameters. Table~\ref{table:implementation} summarizes the main implementation settings, including the 256-bit key size for all cryptographic mechanisms and the use of different algorithms.






\begin{table}[h!]
\centering
    {\scriptsize
    \captionsetup{justification=raggedright,singlelinecheck=false}
    \caption{Implementation details.}
    \label{table:implementation}
    \begin{minipage}{\linewidth}
    \centering
    \begingroup
    \renewcommand{\arraystretch}{1.5}
    \setlength{\tabcolsep}{6pt}
    {
            \begin{tabularx}{\linewidth}{>{\centering\arraybackslash}X >{\centering\arraybackslash}X}
            \hline
            \textbf{Component} & \textbf{Details} \\ 
            \hline
            
            Key length & 256 bits \\
            QKD protocol & BB84$^{\textit{*}}$ \\
            PQC KEM algorithm & ML-KEM-768 \\
            KDF algorithm & HKDF-SHA256 \\

        \end{tabularx}
        }
    \endgroup
    \par\vspace{0.6em}\hrule\par\vspace{0.6em}
    \raggedright
    \textit{$^{*}$Software implementation with SimulaQron simulator.}
    \end{minipage}
    }
\end{table}

\subsection{\textbf{Definition of Key Performance Indicators}}

Based on the QuLore framework implementation and the deployed testbed, to assess the feasibility and performance of QuLore, different KPIs were defined:

\begin{enumerate}
    \item Latency of the E2E key establishment procedure for each Security Level.
    \item Latency of each main subprocess of the framework for each Security Level:
    \begin{enumerate}
        \item Latency of Security Level Assignment.
        \item Latency of Security Level Configuration.
        \item Latency of Secure Key Derivation.
    \end{enumerate}
    \item CPU usage during the E2E key establishment for each Security Level.
    \item Overhead of the communication associated with each Security Level.
\end{enumerate}

The following subsection presents and discusses the experimental results, providing insights into the latency behaviour, CPU consumption, and communication overhead introduced by each process.

\subsection{\textbf{Results and Discussion}}
To evaluate the performance of QuLore, latency measurements were collected for each Security Level. First, complete E2E key establishment across all Security Levels was analyzed, which accounts for all operations required for both the initiating and target applications. The delay associated with transmitting the key identifier from the initiating to the target application was excluded, as it is application-specific and out of the scope of this validation. 

In addition to the overall E2E latency, a detailed analysis was performed for each individual phase, Security Level Assignment, Configuration, and Secure Key Derivation, in order to quantify their individual contributions to total latency. Each test case defined in Table \ref{table:validation} (same as those defined in Figure~\ref{fig:four-with-legend}) was executed 100 times to obtain statistically significant measurements. The generated boxplots have been calculated with 98 percentile and show the median (horizontal line), the mean (x), the box (top limit 75 percentile and bottom limit 25 percentile), and the “whiskers” (that contain the 98\% of results).

Beyond latency, CPU consumption was measured using the \textit{resource} package in Python, which captures the cumulative user and system time consumed by each process during each execution. Communication overhead was assessed by analyzing the communication traces generated for each test case, enabling quantification of the protocol-induced overhead independent of application-level data exchange.



\begin{table}[t]
\centering
{\scriptsize
\captionsetup{justification=raggedright,singlelinecheck=false}
\caption{Test cases}
\label{table:validation}

\begin{minipage}{\linewidth}
\centering
\begingroup
\renewcommand{\arraystretch}{1.5}
\setlength{\tabcolsep}{6pt}
{
\begin{tabularx}{\linewidth}{
  >{\centering\arraybackslash}m{0.12\linewidth}
  >{\centering\arraybackslash}X
  >{\centering\arraybackslash}X
  >{\centering\arraybackslash}X}
\hline
\textbf{Test} & \textbf{Source App} & \textbf{Target App} & \textbf{Security Level} \\
\hline
T1 & APP\_E & APP\_F & Level 1 \\
T2 & APP\_D & APP\_F & Level 2 \\
T3 & APP\_C & APP\_D & Level 3 \\
T4 & APP\_A & APP\_B & Level 4 \\
\hline
\end{tabularx}
}
\endgroup
\end{minipage}
}
\end{table}

\begin{figure}[h!]
\centering
    \includegraphics[width=1\columnwidth]{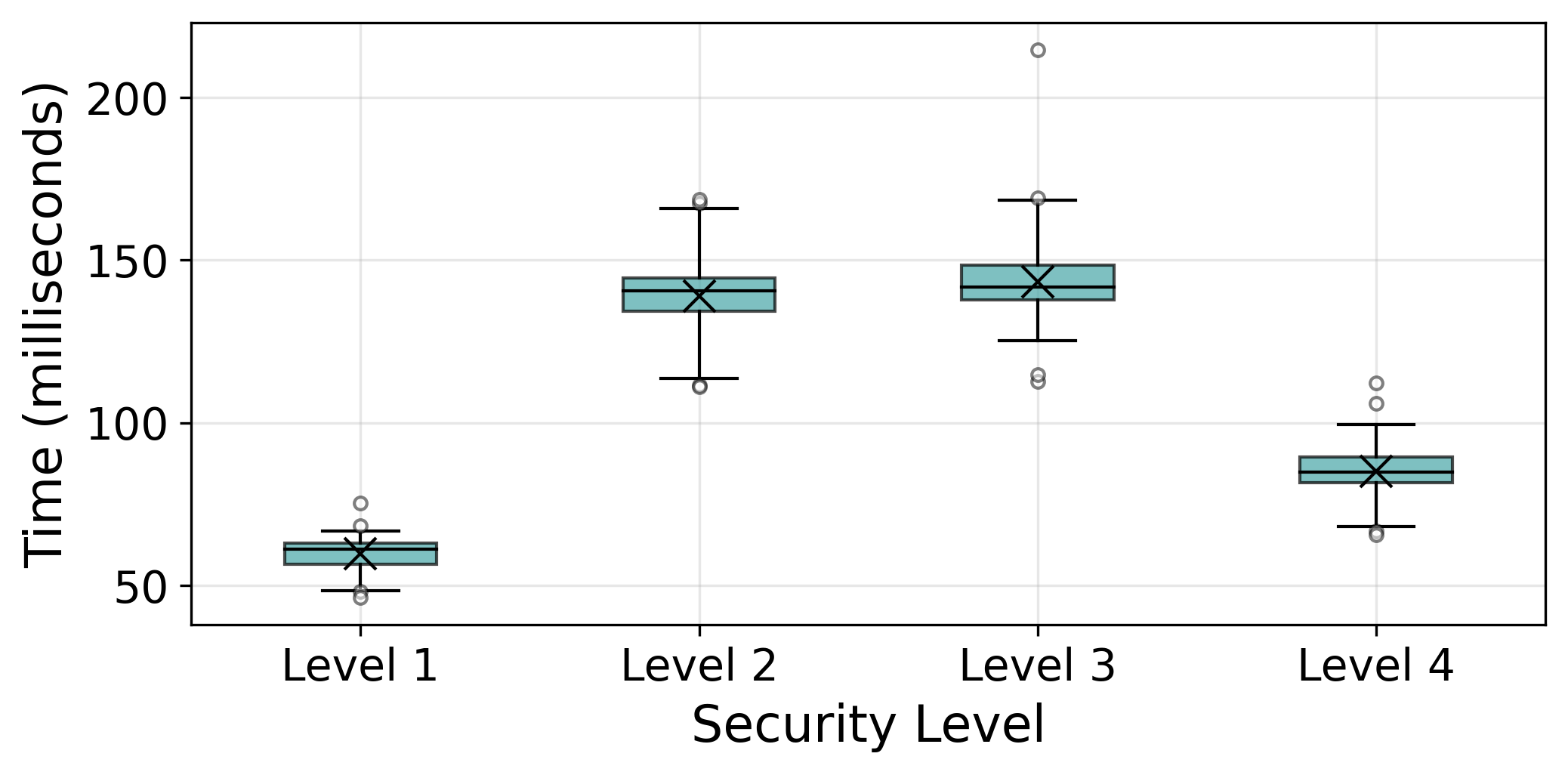}
\caption{Complete E2E key establishment time.}
\label{fig:E2E-time}
\end{figure}

Figure~\ref{fig:E2E-time} shows the total delay for E2E key establishment. Level~1 demonstrates the fastest establishment, averaging 59.81~ms through direct QKD key retrieval, followed by Level~4 with 85.06~ms via purely PQC-based operations. In contrast, the more complex Security Levels exhibited higher latencies, with 138.92~ms for Level~2 and 143.27~ms for Level~3. Additionally, this latency is highly dependent on the number of intermediate trusted nodes that are involved in the key establishmentment. Despite the performance differences across levels, all key establishment operations remained below 150~ms, confirming the framework's suitability.

Moving towards the first subprocess, Security Level Assignment was highly consistent across all scenarios, with assignment times around 6~ms for both initiating and target applications. This uniformity demonstrates that QuSeC performs the decision-making process efficiently, contributing negligible overhead to total E2E time.

\begin{figure}[h!]
\centering
    \includegraphics[width=1\columnwidth]{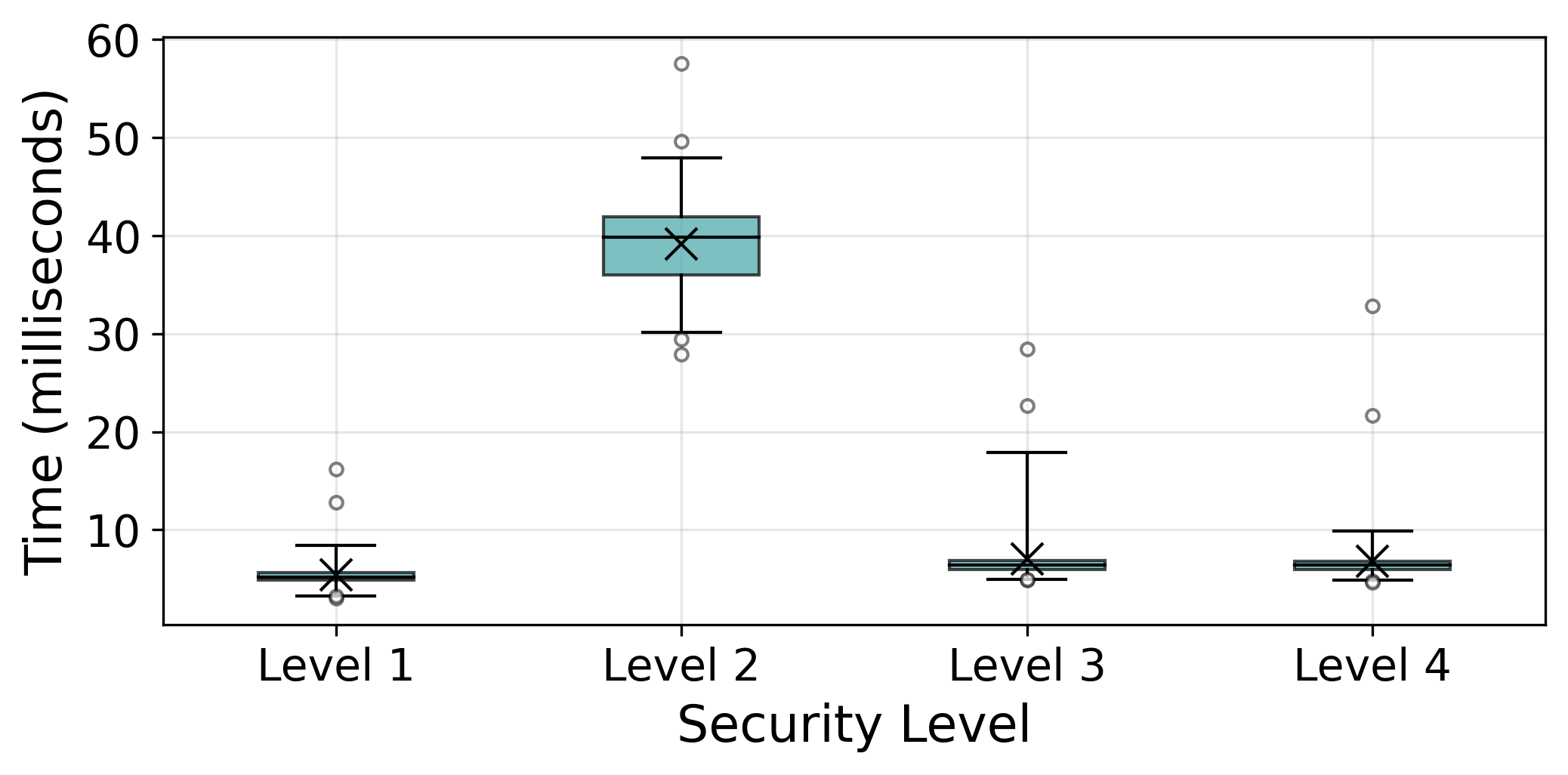}
\caption{Security Level Configuration time for initiating applications.}
\label{fig:config-time}
\end{figure}

The Security Level Configuration phase for the initiating application, shown in Figure~\ref{fig:config-time}, varied more significantly between levels. Level 2 exhibited the highest latency at 39.19~ms, due to the complexity of relay path computation and rule installation across multiple intermediate KMS nodes. This reflects the coordination complexity demands for multi-hop QKD scenarios. The remaining Security Levels maintain low configuration times, with average values ranging from 5 to 7~ms. These results indicate that configuration complexity in the proposed framework is primarily driven by QKD relay infrastructure rather than PQC operations.

\begin{figure}[h!]
\centering
    \includegraphics[width=1\columnwidth]{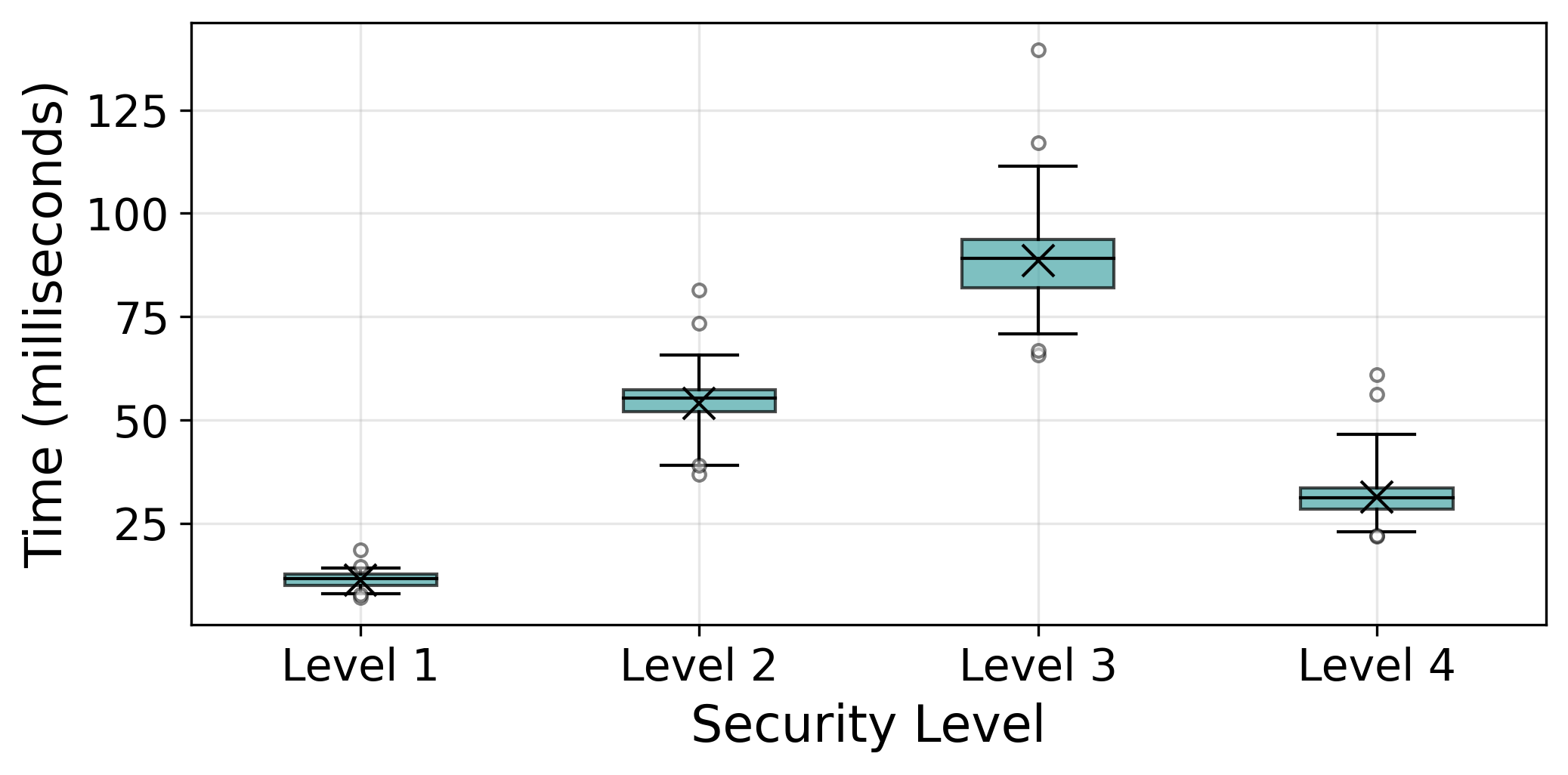}
\caption{Secure Key Derivation time for initiating applications.}
\label{fig:derivation-time}
\end{figure}

The final phase, Secure Key Derivation, showed the largest variation across Security Levels, as depicted in Figure~\ref{fig:derivation-time} for initiating applications. Level 3 exhibited the highest processing time, averaging 88.63~ms, the complexity of the six-step hybrid mechanism that includes PQC KEM establishment, QKD key relay, OTP operations, and KDF processing. Level 2 followed with 54.10~ms, primarily due to multi-hop relay operations and OTP-based key forwarding. Conversely, Level~4 and Level~1 achieved significantly lower derivation times, 31.47~ms and 11.38~ms respectively, owing to their streamlined processes.

\begin{figure}[h!]
\centering
    \includegraphics[width=1\columnwidth]{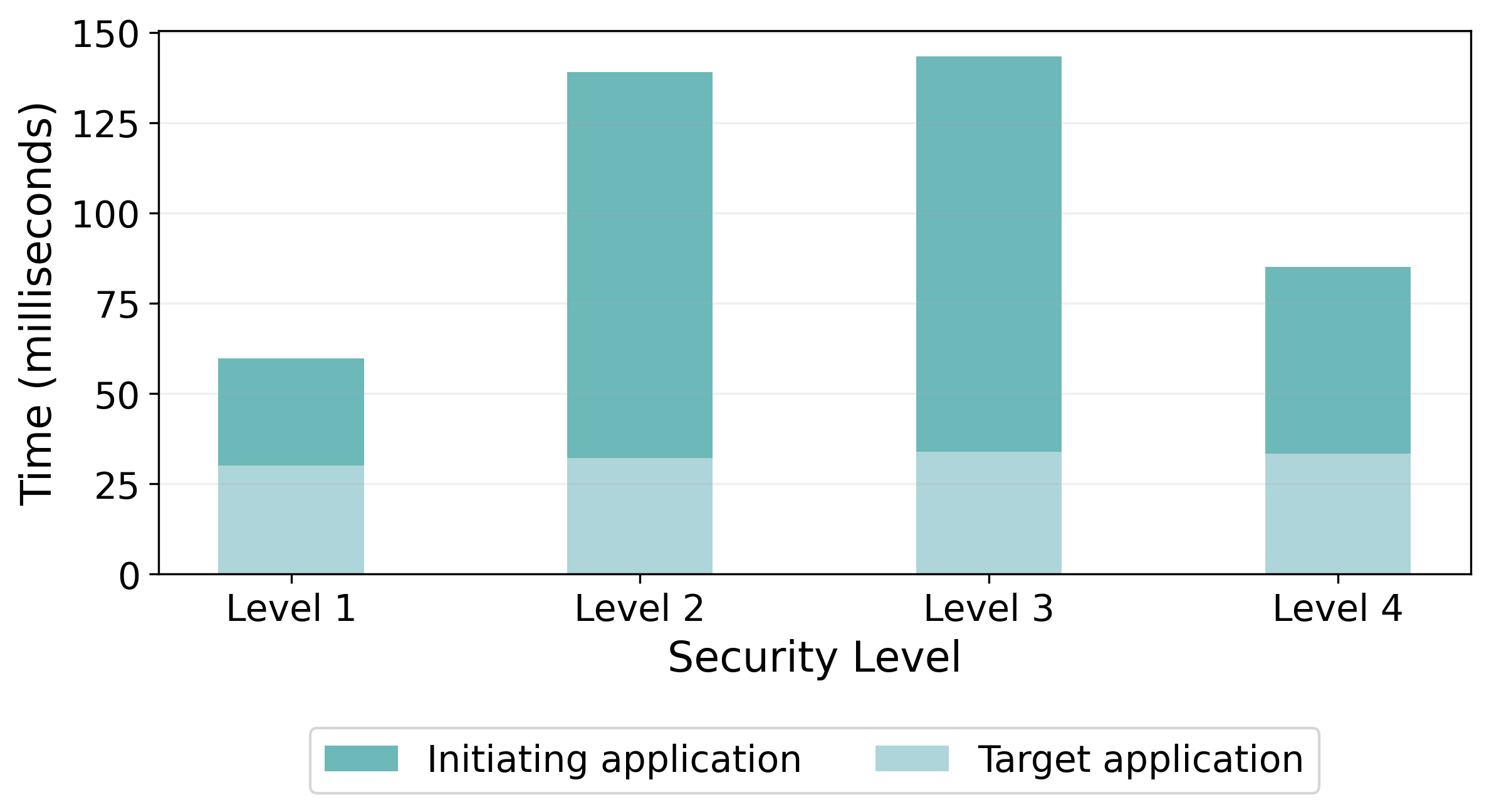}
\caption{Contribution of initiating and target applications to the complete E2E key establishment time.}
\label{fig:init-target}
\end{figure}

Throughout all phases, target application procedures consistently demonstrate lower delays than those of the initiating application. This is a direct result of the QuSeC's session state management, which prevents redundant processing for subsequent endpoints. As represented in Figure~\ref{fig:init-target}, the target application's contribution to total E2E time remains stable across all Security Levels, at approximately 30-35~ms, while the initiating application's contribution varies according to the operations required by each subprocess.

These results validate the framework's principle of adaptive security assignment. Measured latencies demonstrate that enhanced security features, from Level~1's information-theoretic guarantees to Level~3's hybrid quantum-safe protection, can be achieved without excessive operational penalties. This performance evaluation therefore confirms the framework's practical viability for deployment in heterogeneous quantum-safe networks. 

\begin{figure}[h!]
\centering
    \includegraphics[width=0.9\columnwidth]{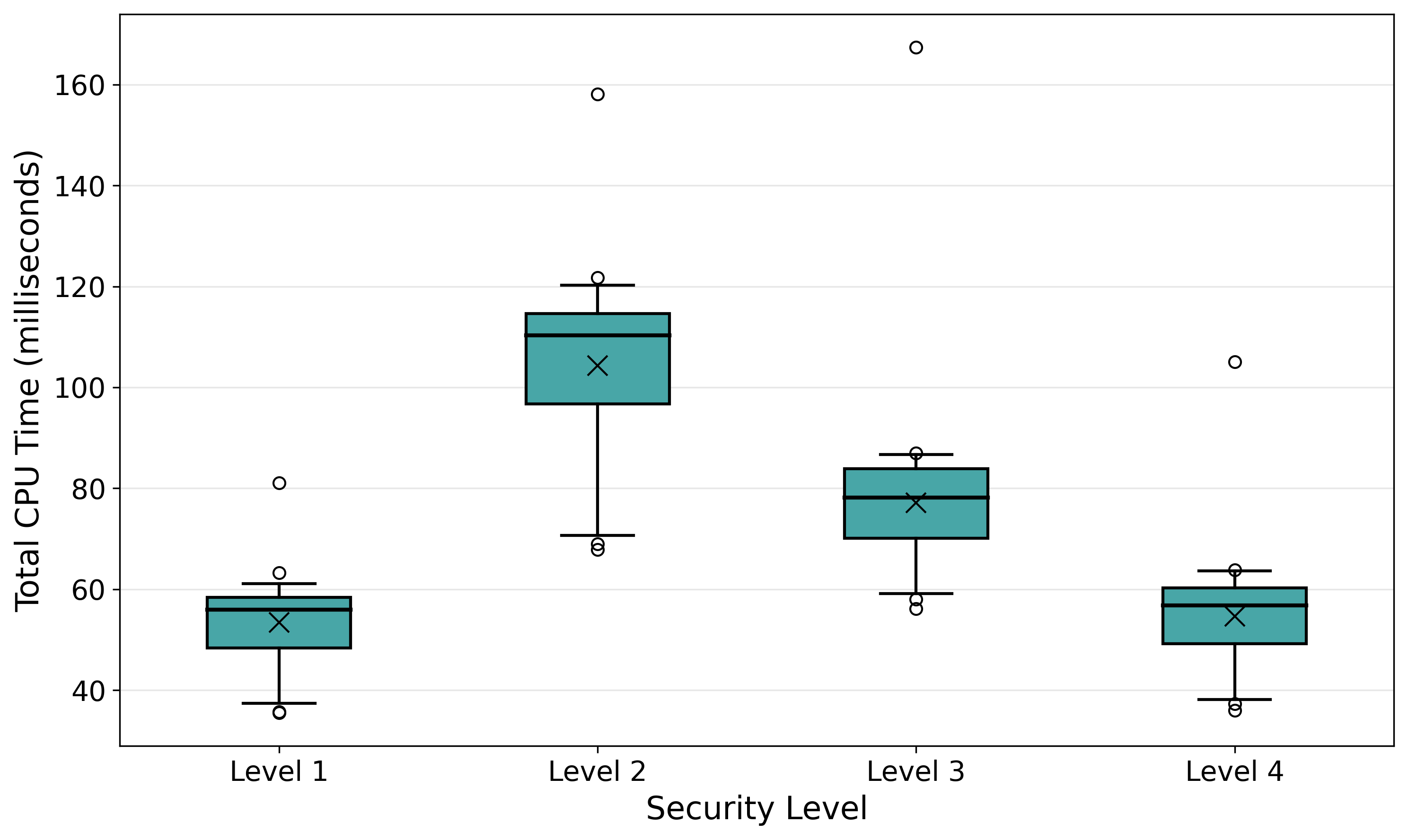}
\caption{Total CPU usage of the E2E key establishment across levels.}
\label{fig:cpu}
\end{figure}

Regarding CPU consumption during E2E key establishment, results show that CPU cost increases with protocol complexity across Security Levels and remains within the sub-120 ms range in the evaluated setup: 53.43 ms (L1), 104.35 ms (L2), 77.10 ms (L3), and 54.66 ms (L4) of total CPU time per establishment, as depicted in Figure~\ref{fig:cpu}.

Beyond absolute CPU time, the ratio between CPU time and E2E latency indicates whether performance is dominated by computation or by distributed coordination. Level 1 is largely CPU-bound (53.43~ms CPU vs 59.81~ms latency, 90\%), reflecting a short workflow with limited messaging. Level 2 remains mostly CPU-bound but with a larger coordination component (104.35~ms CPU vs 138.92~ms latency, 75\%), consistent with additional processing required for trusted-relay handling. In contrast, Level 3 is primarily coordination/communication-bound (77.10~ms CPU vs 143.27~ms latency, ~54\%), since its hybrid procedure introduces multiple message exchanges and waiting periods even though local computation remains moderate. Level 4 shows intermediate behavior (54.66~ms CPU vs 85.06~ms latency, ~65\%), where the remaining gap is attributable to communication and protocol orchestration.

\begin{figure}[h!]
\centering
    \includegraphics[width=1\columnwidth]{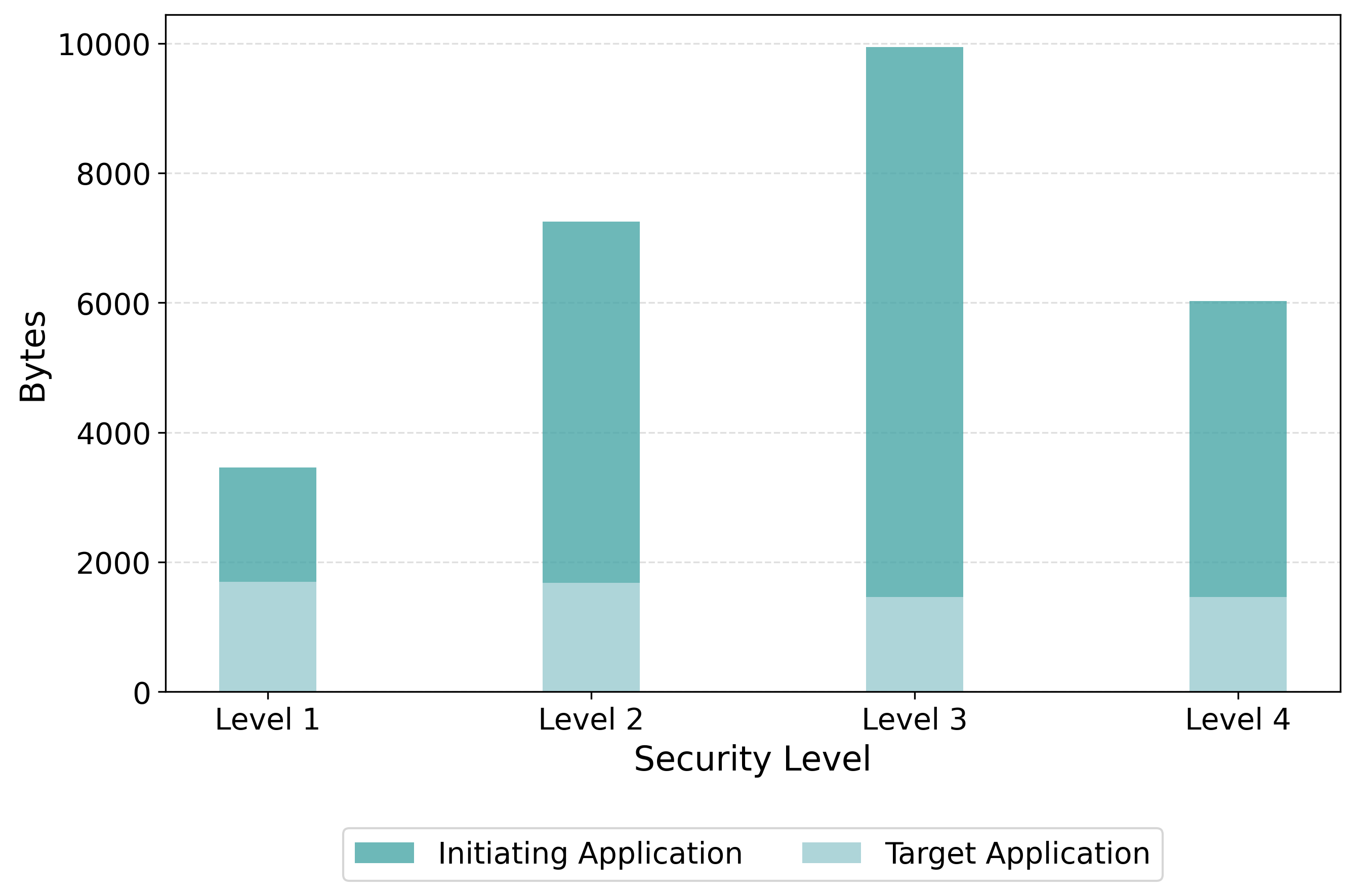}
\caption{Communication overhead of the E2E key establishment across Security Levels.}
\label{fig:overhead}
\end{figure}

Finally, communication overhead was quantified as the total number of bytes exchanged during E2E establishment, with results summarized in Figure~\ref{fig:overhead}. Overhead remains small across all Security Levels: 3.46~kB (L1), 7.25~kB (L2), 9.94~kB (L3), and 6.02~kB (L4). Even the most expensive configuration (L3) stays below 10 kB per key establishment, indicating that the protocol is lightweight in terms of network load.  It is important to note that this overhead remains consistent across all Security Levels, except for Level 2, where additional hops require extra path installation rules and key relay messages.

\section{Conclusions}
\label{sec:conclusions}

This work addresses the need for quantum-safe communications in real-world heterogeneous networks by introducing QuLore: an adaptive framework that integrates QKD and PQC under centralized management. Unlike existing approaches limited to fully QKD-capable environments, this proposal extends quantum protection to all nodes, ensuring robust protection regardless of infrastructure constraints. By defining four dynamic security levels, the system enables tailored protection, balancing information-theoretic guarantees with practical deployability. While the default behavior prioritizes maximum security, the framework can be extended to consider specific service requirements and can even deny service requests when minimum security requirements cannot be satisfied.

The implementation and validation results demonstrate the QuLore framework's ability to deliver secure keys across all communication scenarios in a secure manner. All tested security levels achieved reliable E2E operation, while maintaining full compliance with existing ETSI specifications. Performance evaluation further confirmed the feasibility of the proposed approach, highlighting the potential of the framework for deployment in transitional scenarios where QKD adoption is partial or incremental. Additionally, while the centralized design introduces a potential single point of failure, this risk can be mitigated by deploying QuSeC in a replicated configuration with state synchronization across redundant instances via east–west interfaces. This approach also supports scalability in large deployments by distributing control-plane load without affecting key confidentiality.

In summary, this work establishes a practical path toward hybrid quantum-safe networks. By unifying QKD and PQC into a standard-aware, adaptive model, QuLore provides the flexibility needed to adopt evolving technologies.

\section*{ACKNOWLEDGMENT}
This work was supported in part by the European Commission through the project GN5-2 HORIZON-INFRA-2024-GEANT-01-SGA in the \textit{Working Group 6 Task 1 - Technology}, in part by the Spanish Ministry of Science and Innovation in the project EnablIng Native-AI Secure deterministic 6G networks for hyPer-connected envIRonmEnts (6G-INSPIRE) (PID2022-137329OB-C44), and and in collaboration with the National Hub of Excellence in Quantum Communications, funded by the Ministry of Digital Transition and Public Administration with European Union funds – Next Generation EU, under Component 16.R1 of the Recovery, Transformation, and Resilience Plan.

\bibliographystyle{IEEEtran}
\bibliography{cas-refs}
\vspace{12pt}

\end{document}